%% file: ms.tex
\documentclass[twocolumn]{aastex62}
\usepackage{apjfonts}
\usepackage{natbib}
\usepackage{graphics}
\usepackage{multirow}
\usepackage{amsbsy}
\usepackage{mathrsfs}
\usepackage{footmisc}
\usepackage{amsmath}
\usepackage{mathtools}
\usepackage{amssymb}
\usepackage{comment}
\usepackage{hyperref}
\hypersetup{
  colorlinks = true,
  urlcolor = blue,
  linkcolor=blue,
  citecolor=blue
}
\input{defs}

\newcommand{\ph}{\phantom}

\shorttitle{BK15 Beams and Temperature-to-Polarization Leakage}
\shortauthors{\keckarray\ and \biceptwo\ Collaborations}

\begin{document}

\title{\biceptwo/\keckarray\ XI: Beam Characterization and
  Temperature-to-Polarization Leakage in the BK15 Dataset}

\collaboration{\keckarray\ and \biceptwo\ Collaborations}
\author{P.~A.~R.~Ade}
\affiliation{School of Physics and Astronomy, Cardiff University,
  Cardiff, CF24 3AA, United Kingdom}
\author{Z.~Ahmed}
\affiliation{Kavli Institute for Particle Astrophysics and Cosmology,
  SLAC National Accelerator Laboratory, 2575 Sand Hill Rd, Menlo Park,
  CA 94025, USA}
\author{R.~W.~Aikin}
\affiliation{Department of Physics, California Institute of
  Technology, Pasadena, California 91125, USA}
\author{D.~Barkats}
\affiliation{Harvard-Smithsonian Center for Astrophysics, 60 Garden
  Street MS 42, Cambridge, Massachusetts 02138, USA}
\author{S.~J.~Benton}
\affiliation{Department of Physics, Princeton University, Princeton,
  New Jersey 08544, USA}
\author{C.~A.~Bischoff}
\affiliation{Department of Physics, University of Cincinnati,
  Cincinnati, Ohio 45221, USA}
\author{J.~J.~Bock}
\affiliation{Department of Physics, California Institute of
  Technology, Pasadena, California 91125, USA}
\affiliation{Jet Propulsion Laboratory, Pasadena, California 91109,
  USA}
\author{R.~Bowens-Rubin}
\affiliation{Harvard-Smithsonian Center for Astrophysics, 60 Garden
  Street MS 42, Cambridge, Massachusetts 02138, USA}
\author{J.~A.~Brevik}
\affiliation{Department of Physics, California Institute of
  Technology, Pasadena, California 91125, USA}
\author{I.~Buder}
\affiliation{Harvard-Smithsonian Center for Astrophysics, 60 Garden
  Street MS 42, Cambridge, Massachusetts 02138, USA}
\author{E.~Bullock}
\affiliation{Minnesota Institute for Astrophysics, University of
  Minnesota, Minneapolis, Minnesota 55455, USA}
\author{V.~Buza}
\affiliation{Harvard-Smithsonian Center for Astrophysics, 60 Garden
  Street MS 42, Cambridge, Massachusetts 02138, USA}
\affiliation{Department of Physics, Harvard University, Cambridge, MA
  02138, USA}
\author{J.~Connors}
\affiliation{Harvard-Smithsonian Center for Astrophysics, 60 Garden
  Street MS 42, Cambridge, Massachusetts 02138, USA}
\author{J.~Cornelison}
\affiliation{Harvard-Smithsonian Center for Astrophysics, 60 Garden
  Street MS 42, Cambridge, Massachusetts 02138, USA}
\author{B.~P.~Crill}
\affiliation{Jet Propulsion Laboratory, Pasadena, California 91109,
  USA}
\author{M.~Crumrine}
\affiliation{School of Physics and Astronomy, University of Minnesota,
  Minneapolis, Minnesota 55455, USA}
\author{M.~Dierickx}
\affiliation{Harvard-Smithsonian Center for Astrophysics, 60 Garden
  Street MS 42, Cambridge, Massachusetts 02138, USA}
\author{L.~Duband}
\affiliation{Service des Basses Temp\'{e}ratures, Commissariat \`{a}
  l'Energie Atomique, 38054 Grenoble, France}
\author{J.~P.~Filippini}
\affiliation{Department of Physics, University of Illinois at
  Urbana-Champaign, Urbana, Illinois 61801, USA}
\affiliation{Department of Astronomy, University of Illinois at
  Urbana-Champaign, Urbana, Illinois 61801, USA}
\author{S.~Fliescher}
\affiliation{School of Physics and Astronomy, University of Minnesota,
  Minneapolis, Minnesota 55455, USA}
\author{J.~Grayson}
\affiliation{Department of Physics, Stanford University, Stanford, CA
  94305, USA}
\author{G.~Hall}
\affiliation{School of Physics and Astronomy, University of Minnesota,
  Minneapolis, Minnesota 55455, USA}
\author{M.~Halpern}
\affiliation{Department of Physics and Astronomy, University of
  British Columbia, Vancouver, British Columbia, V6T 1Z1, Canada}
\author{S.~Harrison}
\affiliation{Harvard-Smithsonian Center for Astrophysics, 60 Garden
  Street MS 42, Cambridge, Massachusetts 02138, USA}
\author{S.~R.~Hildebrandt}
\affiliation{Department of Physics, California Institute of
  Technology, Pasadena, California 91125, USA}
\affiliation{Jet Propulsion Laboratory, Pasadena, California 91109,
  USA}
\author{G.~C.~Hilton}
\affiliation{National Institute of Standards and Technology, Boulder,
  Colorado 80305, USA}
\author{H.~Hui}
\affiliation{Department of Physics, California Institute of
  Technology, Pasadena, California 91125, USA}
\author{K.~D.~Irwin}
\affiliation{Department of Physics, Stanford University, Stanford, CA
  94305, USA}
\affiliation{Kavli Institute for Particle Astrophysics and Cosmology,
  SLAC National Accelerator Laboratory, 2575 Sand Hill Rd, Menlo Park,
  CA 94025, USA}
\affiliation{National Institute of Standards and Technology, Boulder,
  Colorado 80305, USA}
\author{J.~Kang}
\affiliation{Department of Physics, Stanford University, Stanford, CA
  94305, USA}
\author{K.~S.~Karkare}
\affiliation{Harvard-Smithsonian Center for Astrophysics, 60 Garden
  Street MS 42, Cambridge, Massachusetts 02138, USA}
\affiliation{Kavli Institute for Cosmological Physics, University of
  Chicago, Chicago, IL 60637, USA}
\author{E.~Karpel}
\affiliation{Department of Physics, Stanford University, Stanford, CA
  94305, USA}
\author{J.~P.~Kaufman}
\affiliation{Department of Physics, University of California at San
  Diego, La Jolla, California 92093, USA}
\author{B.~G.~Keating}
\affiliation{Department of Physics, University of California at San
  Diego, La Jolla, California 92093, USA}
\author{S.~Kefeli}
\affiliation{Department of Physics, California Institute of
  Technology, Pasadena, California 91125, USA}
\author{S.~A.~Kernasovskiy}
\affiliation{Department of Physics, Stanford University, Stanford, CA
  94305, USA}
\author{J.~M.~Kovac}
\affiliation{Harvard-Smithsonian Center for Astrophysics, 60 Garden
  Street MS 42, Cambridge, Massachusetts 02138, USA}
\affiliation{Department of Physics, Harvard University, Cambridge, MA
  02138, USA}
\author{C.~L.~Kuo}
\affiliation{Department of Physics, Stanford University, Stanford, CA
  94305, USA}
\affiliation{Kavli Institute for Particle Astrophysics and Cosmology,
  SLAC National Accelerator Laboratory, 2575 Sand Hill Rd, Menlo Park,
  CA 94025, USA}
\author{N.~A.~Larsen}
\affiliation{Kavli Institute for Cosmological Physics, University of
  Chicago, Chicago, IL 60637, USA}
\author{K.~Lau}
\affiliation{School of Physics and Astronomy, University of Minnesota,
  Minneapolis, Minnesota 55455, USA}
\author{E.~M.~Leitch}
\affiliation{Kavli Institute for Cosmological Physics, University of
  Chicago, Chicago, IL 60637, USA}
\author{M.~Lueker}
\affiliation{Department of Physics, California Institute of
  Technology, Pasadena, California 91125, USA}
\author{K.~G.~Megerian}
\affiliation{Jet Propulsion Laboratory, Pasadena, California 91109,
  USA}
\author{L.~Moncelsi}
\affiliation{Department of Physics, California Institute of
  Technology, Pasadena, California 91125, USA}
\author{T.~Namikawa}
\affiliation{Department of Applied Mathematics and Theoretical
  Physics, University of Cambridge, Cambridge, CB3 0WA, United
  Kingdom}
\author{C.~B.~Netterfield}
\affiliation{Department of Physics, University of Toronto, Toronto,
  Ontario, M5S 1A7, Canada}
\affiliation{Canadian Institute for Advanced Research, Toronto,
  Ontario, M5G 1Z8, Canada}
\author{H.~T.~Nguyen}
\affiliation{Jet Propulsion Laboratory, Pasadena, California 91109,
  USA}
\author{R.~O'Brient}
\affiliation{Department of Physics, California Institute of
  Technology, Pasadena, California 91125, USA}
\affiliation{Jet Propulsion Laboratory, Pasadena, California 91109,
  USA}
\author{R.~W.~Ogburn~IV}
\affiliation{Department of Physics, Stanford University, Stanford, CA
  94305, USA}
\affiliation{Kavli Institute for Particle Astrophysics and Cosmology,
  SLAC National Accelerator Laboratory, 2575 Sand Hill Rd, Menlo Park,
  CA 94025, USA}
\author{S.~Palladino}
\affiliation{Department of Physics, University of Cincinnati,
  Cincinnati, Ohio 45221, USA}
\author{C.~Pryke}
\affiliation{School of Physics and Astronomy, University of Minnesota,
  Minneapolis, Minnesota 55455, USA}
\affiliation{Minnesota Institute for Astrophysics, University of
  Minnesota, Minneapolis, Minnesota 55455, USA}
\author{B.~Racine}
\affiliation{Harvard-Smithsonian Center for Astrophysics, 60 Garden
  Street MS 42, Cambridge, Massachusetts 02138, USA}
\author{S.~Richter}
\affiliation{Harvard-Smithsonian Center for Astrophysics, 60 Garden
  Street MS 42, Cambridge, Massachusetts 02138, USA}
\author{A.~Schillaci}
\affiliation{Department of Physics, California Institute of
  Technology, Pasadena, California 91125, USA}
\author{R.~Schwarz}
\affiliation{School of Physics and Astronomy, University of Minnesota,
  Minneapolis, Minnesota 55455, USA}
\author{C.~D.~Sheehy}
\affiliation{Physics Department, Brookhaven National Laboratory,
  Upton, New York 11973, USA}
\author{A.~Soliman}
\affiliation{Department of Physics, California Institute of
  Technology, Pasadena, California 91125, USA}
\author{T.~St.~Germaine}
\affiliation{Harvard-Smithsonian Center for Astrophysics, 60 Garden
  Street MS 42, Cambridge, Massachusetts 02138, USA}
\author{Z.~K.~Staniszewski}
\affiliation{Department of Physics, California Institute of
  Technology, Pasadena, California 91125, USA}
\affiliation{Jet Propulsion Laboratory, Pasadena, California 91109,
  USA}
\author{B.~Steinbach}
\affiliation{Department of Physics, California Institute of
  Technology, Pasadena, California 91125, USA}
\author{R.~V.~Sudiwala}
\affiliation{School of Physics and Astronomy, Cardiff University,
  Cardiff, CF24 3AA, United Kingdom}
\author{G.~P.~Teply}
\affiliation{Department of Physics, California Institute of
  Technology, Pasadena, California 91125, USA}
\affiliation{Department of Physics, University of California at San
  Diego, La Jolla, California 92093, USA}
\author{K.~L.~Thompson}
\affiliation{Department of Physics, Stanford University, Stanford, CA
  94305, USA}
\affiliation{Kavli Institute for Particle Astrophysics and Cosmology,
  SLAC National Accelerator Laboratory, 2575 Sand Hill Rd, Menlo Park,
  CA 94025, USA}
\author{J.~E.~Tolan}
\affiliation{Department of Physics, Stanford University, Stanford, CA
  94305, USA}
\author{C.~Tucker}
\affiliation{School of Physics and Astronomy, Cardiff University,
  Cardiff, CF24 3AA, United Kingdom}
\author{A.~D.~Turner}
\affiliation{Jet Propulsion Laboratory, Pasadena, California 91109,
  USA}
\author{C.~Umilta}
\affiliation{Department of Physics, University of Cincinnati,
  Cincinnati, Ohio 45221, USA}
\author{A.~G.~Vieregg}
\affiliation{Department of Physics, Enrico Fermi Institute, University
  of Chicago, Chicago, IL 60637, USA}
\affiliation{Kavli Institute for Cosmological Physics, University of
  Chicago, Chicago, IL 60637, USA}
\author{A.~Wandui}
\affiliation{Department of Physics, California Institute of
  Technology, Pasadena, California 91125, USA}
\author{A.~C.~Weber}
\affiliation{Jet Propulsion Laboratory, Pasadena, California 91109,
  USA}
\author{D.~V.~Wiebe}
\affiliation{Department of Physics and Astronomy, University of
  British Columbia, Vancouver, British Columbia, V6T 1Z1, Canada}
\author{J.~Willmert}
\affiliation{School of Physics and Astronomy, University of Minnesota,
  Minneapolis, Minnesota 55455, USA}
\author{C.~L.~Wong}
\affiliation{Harvard-Smithsonian Center for Astrophysics, 60 Garden
  Street MS 42, Cambridge, Massachusetts 02138, USA}
\affiliation{Department of Physics, Harvard University, Cambridge, MA
  02138, USA}
\author{W.~L.~K.~Wu}
\affiliation{Kavli Institute for Cosmological Physics, University of
  Chicago, Chicago, IL 60637, USA}
\author{H.~Yang}
\affiliation{Department of Physics, Stanford University, Stanford, CA
  94305, USA}
\author{K.~W.~Yoon}
\affiliation{Department of Physics, Stanford University, Stanford, CA
  94305, USA}
\affiliation{Kavli Institute for Particle Astrophysics and Cosmology,
  SLAC National Accelerator Laboratory, 2575 Sand Hill Rd, Menlo Park,
  CA 94025, USA}
\author{C.~Zhang}
\affiliation{Department of Physics, California Institute of
  Technology, Pasadena, California 91125, USA}

\correspondingauthor{K.~S.~Karkare}
\email{kkarkare@kicp.uchicago.edu}

\begin{abstract}
Precision measurements of cosmic microwave background (CMB)
polarization require extreme control of instrumental systematics.  In
a companion paper we have presented cosmological constraints from
observations with the \biceptwo\ and \keckarray\ experiments up to and
including the 2015 observing season (BK15), resulting in the deepest
CMB polarization maps to date and a statistical sensitivity to the
tensor-to-scalar ratio of $\sigma(r) = 0.020$.  In this work we
characterize the beams and constrain potential systematic
contamination from main beam shape mismatch at the three BK15
frequencies (95, 150, and 220\,GHz).  Far-field maps of 7,360 distinct
beam patterns taken from 2010--2015 are used to measure differential
beam parameters and predict the contribution of
temperature-to-polarization leakage to the BK15 \bmode\ maps.  In the
multifrequency, multicomponent likelihood analysis that uses BK15,
\planck, and \wmap\ maps to separate sky components, we find that
adding this predicted leakage to simulations induces a bias of $\Delta
r = 0.0027 \pm 0.0019$.  Future results using higher-quality beam maps
and improved techniques to detect such leakage in CMB data will
substantially reduce this uncertainty, enabling the levels of
systematics control needed for \bicep\ Array and other experiments
that plan to definitively probe large-field inflation.
\end{abstract}


\section{Introduction}
\label{sec:intro}

Progress in understanding the physics of the early Universe through
measurement of the cosmic microwave background (CMB) has been driven
by advances in instrumental sensitivity.  Since its discovery over 50
years ago \citep{penzias65}, successively finer spatial features in
the CMB have been detected, including the 3\,mK dipole
\citep{conklin69}, the $\sim100$\,$\mu$K temperature fluctuations
\citep{smoot92}, the $\sim1$\,$\mu$K $E$-mode polarization
anisotropies \citep{kovac02}, and most recently the $\sim100$\,nK
$B$-mode polarization from gravitational lensing of $E$ modes
\citep{hanson13, polarbear14, keisler15, louis17, biceptwoVIII,
  planck_lensing}.  Confidence in these measurements requires
constraining the effects of instrumental systematics to well below the
statistical uncertainty, which is determined by factors such as
detector count, scan strategy, and astrophysical component separation.

One of the next frontiers in CMB measurements is constraining
degree-scale $B$-mode polarization that may have been imprinted on the
surface of last scattering by gravitational waves---a distinctive
feature of inflationary models \citep{kamionkowski06}.  Such a
measurement is made challenging by the unknown (or potentially
vanishing) amplitude of this signal, parametrized by the
tensor-to-scalar ratio $r$, and the fact that gravitational lensing
and Galactic foregrounds also generate $B$ modes; see \citet{cmbs4}
for a comprehensive review.  The most stringent constraint to date,
presented in the companion paper \cite{biceptwoX} (hereafter BK-X),
uses deep \biceptwo\ and \keckarray\ maps at 95, 150, and 220\,GHz up
to and including the 2015 season (hereafter denoted the BK15 dataset)
in conjunction with external maps from the
\planck\ \citep{planck2015I} and \wmap\ \citep{bennett13} satellites
to separate the CMB from foregrounds.  The resulting $BB$ power
spectra are consistent with a combination of the expected lensing
signal and Galactic dust, yielding a 95\% upper limit of $r_{0.05} <
0.072$ which tightens to $r_{0.05} < 0.062$ when including CMB
temperature and additional data.  The total experimental statistical
sensitivity is $\sigma(r) = 0.020$, which takes into account
uncertainty in foreground separation.

\begin{table*}
\begin{center}
\caption{\biceptwo/\keckarray\ configuration 2010--2015}
\begin{tabular}{|c|c|c|c|c|c|} 
\hline
\rule[-1ex]{0pt}{3.5ex}Rx  & \biceptwo\ 2010--2012 & \keck\ 2012 &\keck\ 2013 &  \keck\ 2014 & \keck\ 2015 \\
\hline
0 & 150 GHz, 512 (432) & 150 GHz, 512 (326) & 150 GHz, 512 (318)  & \ph{0}95 GHz, 288 (224) & \ph{0}95 GHz, 288 (218) \\
1 & & 150 GHz, 512 (408) & 150 GHz, 512 (400)  & 150 GHz, 512 (400) & 220 GHz, 512 (346) \\
2 & & 150 GHz, 512 (314) & 150 GHz, 512 (312)  & \ph{0}95 GHz, 288 (248) & \ph{0}95 GHz, 288 (248) \\
3 & & 150 GHz, 512 (340) & 150 GHz, 512 (422)  & 150 GHz, 512 (398) & 220 GHz, 512 (376) \\
4 & & 150 GHz, 512 (392) & 150 GHz, 512 (386)  & 150 GHz, 512 (388)  & 150 GHz, 512 (378) \\
  \hline
\end{tabular}
\tablecomments{Center frequency, nominal detector count, and detector
  count used in cosmological analysis (in parentheses) for all
  receivers contributing to the BK15 maps.  The nominal
  detector count includes a small number of dark detectors that are
  intentionally disconnected from the antenna (used to characterize
  sensitivity to temperature fluctuations and RF interference) and
  detectors lost to imperfect yield.  The detector count contributing
  to the final maps includes only detectors that have passed all data
  quality cuts.}
\label{tab:config}
\end{center}
\end{table*}

The \biceptwo/\keckarray\ CMB experiments, located at the
Amundsen-Scott South Pole Station, are small-aperture refracting
telescopes which measure polarization by differencing pairs of
co-located orthogonally polarized detectors, resulting in extremely
effective suppression of common-mode noise.  The most prominent
systematic in the measurement is leakage of the bright temperature sky
into polarization ($T \rightarrow P$) due to mismatched beam shapes
within a polarization pair.  In this work, we report on optical
characterization of the \keckarray\ receivers during the 2014 and 2015
observing seasons operating at 95, 150, and 220\,GHz.  We present
high-fidelity far-field beam maps from which we measure Gaussian beam
parameters and validate the ``deprojection'' procedure used to
marginalize over the lowest-order main beam difference modes that
produce the majority of $T \rightarrow P$ leakage.  We then use these
maps in specialized ``beam map simulations'' to derive upper limits on
the higher-order undeprojected $T \rightarrow P$ leakage, and take
cross spectra with the BK15 maps to estimate the systematic
contribution to the real data.  When this leakage is propagated
through the BK15 multicomponent analysis, we recover a bias on $r$
that is subdominant to the statistical sensitivity.  The analysis
techniques explored here offer an example of how we may treat
systematics in future experiments with an order of magnitude greater
sensitivity.

This paper is one in a series of publications by the
\biceptwo/\keckarray\ collaborations and accompanies the primary BK15
CMB results shown in BK-X.  An overview of the
\biceptwo/\keckarray\ instruments is provided in \cite{biceptwoII} and
optical characterization of the array through 2013 is presented in
\cite{biceptwoIV} (hereafter BK-IV).  Here we extend the beam
characterization in BK-IV to the 2014 and 2015 observing seasons, in
which we reconfigured several receivers to operate at frequencies
other than 150\,GHz.  Table~\ref{tab:config} shows the configuration
of \biceptwo/\keckarray\ through 2015.  The beam map simulations and
constraints on $T \rightarrow P$ leakage presented here are based on
those shown in \cite{biceptwoIII} (hereafter BK-III) and include all
beams contributing to the BK15 maps.  In this work the technique is
extended to explicitly test for leakage in the CMB maps and to account
for the potential systematic contribution in multifrequency component
separation.

We organize this paper as follows.  In Section~\ref{sec:ffbm} we
review the setup at South Pole used to measure the far-field beam
pattern of every detector contributing to the CMB maps.  Differential
Gaussian parameters extracted from these maps are presented in
Section~\ref{sec:gauss}.  Composite maps and array-averaged beam
profiles are shown in Section~\ref{sec:composites}.  We then use the
composite beam maps to predict the level of undeprojected $T
\rightarrow P$ leakage in the BK15 dataset in Section~\ref{sec:ttop},
and cross-correlate these predictions with the CMB maps.  We analyze
the impact of $T \rightarrow P$ leakage on the multicomponent
likelihood analysis in Section~\ref{sec:likelihood}, and conclude in
Section~\ref{sec:conclusion}.

\section{Far-Field Beam Measurements}
\label{sec:ffbm}

\begin{figure*}
\begin{center}
\resizebox{1.0\textwidth}{!}{\includegraphics{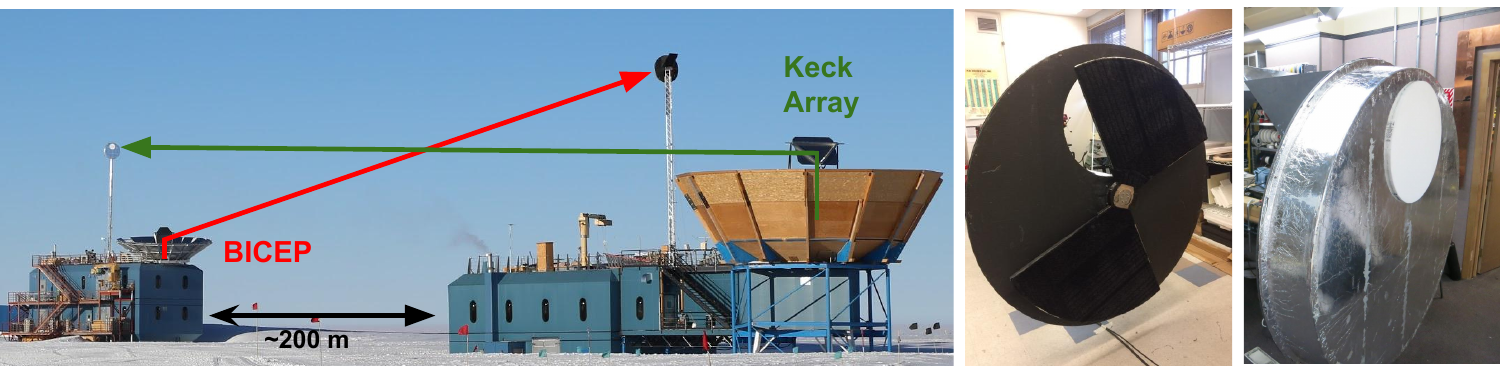}}
\end{center}
\caption{Several weeks of each South Pole deployment season are spent
  generating far-field beam maps using the pictured setup.
  \emph{Left}: \keckarray\ and \bicepthree\ taking far-field beam
  maps simultaneously at South Pole. Both choppers and far-field flat
  mirrors are visible.  \emph{Middle}: Carbon fiber chopper blade
  coated in Eccosorb HR-10, mounted in the lab; during operation it
  spins at 14\,Hz.  \emph{Right}: Carbon fiber enclosure holding the
  blade; the 60 cm aperture (white) is sealed with Zotefoam HD30.
  When the blade does not fill the aperture, the beam is redirected to
  the zenith with a $45^{\circ}$ mirror.  }
\label{fig:chopper_mast}
\end{figure*}

Precision measurements of the \bicep/\keckarray\ beams in the far
field are enabled by our small-aperture approach.  The standard
far-field distance criterion of $2D^2/\lambda$, where $D$ is the
aperture size (26.4\,cm) and $\lambda$ is the wavelength, yields 46,
70, and 103\,m for \keckarray\ receivers at 95, 150, and 220\,GHz
respectively.  Since the Martin A. Pomerantz Observatory (where
\keckarray\ is located) and the Dark Sector Laboratory (where
\biceptwo\ was located) are 210\,m apart, a source mounted on either
building is comfortably in the far field of a receiver on the opposite
building as illustrated in Figure~\ref{fig:chopper_mast}---see
Section~\ref{ss:bmsyst} for more details.

To view the source we install a $45^{\circ}$ flat mirror above the
mount to redirect the beams over the ground shield; when the telescope
is at zenith, the beams point towards the horizon.  The five
\keckarray\ receivers are spread out over 1.5\,m and redirecting all
beams simultaneously would require a larger mirror than can be
feasibly supported by the mount.  Instead the $1.8 \times 2.7$\,m
aluminum honeycomb mirror is mounted so that at any instant several
receivers are completely underneath the mirror.  Rotation about the
boresight axis then allows all receivers to be measured.  Since it is
valuable to measure each receiver at many boresight angles, we move
the mirror to various positions above the mount so that different
angles are accessible (see Figure~\ref{fig:keck_beam_coords}).  The
co-moving absorptive forebaffles are removed so that the mirror can be
mounted.

The source consists of a blade coated with Eccosorb HR-10 microwave
absorber within an enclosure.  As the blade spins, a beam pointed at
the circular aperture on the enclosure alternately sees a hot load
(the $\sim250$\,K ambient-temperature blade) or a cold load (a flat
mirror behind the blade which redirects the beam up to the $\sim12$\,K
sky).  An optical encoder on the rotation axis of the blade is
recorded and used to demodulate the detector timestreams so that only
variation at the chop frequency and phase is interpreted as signal.
In general a larger aperture offers higher signal and therefore faster
mapping speed.  Because the telescope scans in azimuth continuously
while the chopper spins, a faster chop rate reduces systematic
contribution from bright azimuth-fixed signals.  In previous
publications, beam maps were generated with 30 or 45\,cm aperture
choppers that spun at 18 and 10\,Hz chop rates, respectively.  For the
2015--2016 deployment season we used a chopper made with a carbon
fiber composite to reduce the weight (to 55\,lb), featuring a 60\,cm
aperture and spinning at a 14\,Hz chop rate \citep{karkare16}.  The
aperture is closed with Zotefoam HD30 to prevent wind from affecting
the motion.  Figure~\ref{fig:chopper_mast} shows two choppers mounted
on masts at South Pole and in the lab.

Far-field beam maps are generated by scanning across the source in
azimuth and stepping in elevation.  A typical beam measurement spans
$22.8^{\circ}$ in azimuth and $20^{\circ}$ in elevation with steps of
$0.05^{\circ}$, takes $\sim 8$ h, and---given the $\sim 16^{\circ}$
field of view---allows for measurement of all detectors in at least
one \keckarray\ focal plane to $>2^{\circ}$ from each beam center.
The specifics of the beam measurement are determined by factors such
as chop rate, desired map pixelization, and sub-kelvin refrigerator
hold time.  Each year several weeks are spent taking beam map data
before CMB scanning begins.  Since we mask out parts of the map with
known ground-fixed contamination (see Section~\ref{ss:composite_gen}),
maps are made at multiple boresight angles to allow measurement of all
regions of the beam.  Using several boresight angles also enables
rigorous systematics checks.  In a single year 40--50 of the scans
described above are typical, spread out over 10 boresight angles.

The new measurements presented in this paper took place in February
and March of 2014 and 2015; measurements from previous beam mapping
campaigns are also included in the simulation results presented in
Section~\ref{sec:ttop}.  From 2010 to 2015 we measured 10,368 beam
patterns to high precision for both \biceptwo\ and \keckarray, of
which 3,008 were repeats in which the instrument was unchanged from a
previous season, allowing for consistency tests.\footnote{These
  numbers do not reflect the total number of fielded detectors: in one
  year we installed additional baffling inside the cryostats, which
  affected the beam patterns of existing detectors \citep{buder14}.}

\section{Gaussian Beam Parameters}
\label{sec:gauss}

\bicep/\keckarray\ beam shapes are well-approximated by elliptical
Gaussians.  In BK-IV we presented Gaussian parameters for
\biceptwo\ and \keckarray\ from 2010--2013, through which all
receivers operated at 150\,GHz.  For the 2014 observing season two
150\,GHz receivers were converted to 95\,GHz, and for 2015 two more
were converted to 220\,GHz (Table~\ref{tab:config}).  Here we present
beam parameters for 2014 and 2015.

\subsection{Coordinate System}

\begin{figure}
\begin{center}
\resizebox{0.95\columnwidth}{!}{\includegraphics{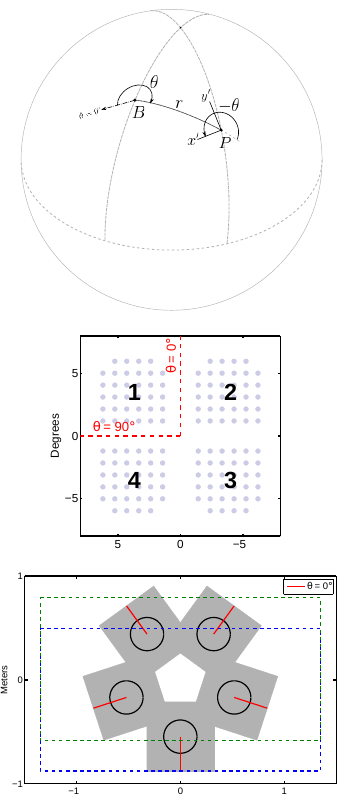}}
\end{center}
\caption{\emph{Top}: The $(x',y')$ coordinate system is centered
  locally for each pixel $P$ at a location $(r,\theta)$ from the
  boresight $B$, and is referenced to the $\theta=0^{\circ}$ ray.
  \emph{Middle}: The $\theta=0^{\circ}$ ray is referenced to the
  orientation of detector tiles, here depicted as projected onto the
  sky as viewed from Earth.  \emph{Bottom}: Schematic of the five
  \keck\ receiver apertures in the mount, clocked at $72^{\circ}$ with
  respect to each other (black circles).  Two footprints of the
  far-field flat mirror (hoisted 3.7\,m above the apertures) are
  depicted as dashed boxes.  The rough extent to which the beam
  centers have diverged (gray boxes) and each receiver's
  $\theta=0^{\circ}$ ray (red lines) are depicted as they intercept
  the mirror.  Multiple mirror positions are necessary to reflect
  beams from all receivers at many boresight angles.}
\label{fig:keck_beam_coords}
\end{figure}

We begin by defining an instrument-fixed spherical coordinate system
which is independent of the orientation of the instrument with respect
to celestial coordinates, illustrated in
Figure~\ref{fig:keck_beam_coords}.  A spatial pixel $P$ containing two
orthogonally polarized detectors is defined to be at a location
$(r,\theta)$ from the boresight $B$, where $r$ is the radial distance
from $B$ and $\theta$ is the counterclockwise angle looking out from
the telescope towards the sky from the $\theta = 0^{\circ}$ ray.  The
$\theta=0^{\circ}$ ray is defined with the choice of a fixed index
angle on the instrument; we choose this to be, from the boresight,
along Tiles 1 and 2 on the focal plane.  Tiles are numbered
counterclockwise looking directly down on the focal plane, and Tiles 1
and 2 are physically located on the side of the focal plane connecting
the heat straps to the sub-kelvin refrigerator.  Each receiver's
$\theta=0^{\circ}$ ray is different, depending on the clocking of the
receiver in the mount.

For each pixel we then define a local $(x',y')$ Cartesian coordinate
system, where the positive $x'$ axis is defined to be along the great
circle passing through the pixel center that is an angle $-\theta$
from the $\hat{r}$ direction of the pixel.  The $y'$ axis is defined
to be the great circle that is $+90^{\circ}$ away from the $x'$ axis.
This $(x',y')$ coordinate system is then projected onto a plane at the
pixel center, and rotates with the instrument on the sky.

\subsection{Beam Fitting and Statistics}

Each measured beam is fit to a two-dimensional elliptical Gaussian
$B(\mathbf{x})$ with six free parameters:
\begin{equation}
B(\mathbf{x}) = \frac{1}{\Omega}e^{-\frac{1}{2}\left(\mathbf{x} - \boldsymbol{\mu} \right)^T \boldsymbol{\Sigma}^{-1} \left(\mathbf{x} - \boldsymbol{\mu} \right)},
\end{equation}
where $\mathbf{x} = (x',y')$ is the beam map coordinate,
$\boldsymbol{\mu} = (x_0,y_0)$ is the location of the beam center,
$\Omega$ is the normalization, and $\boldsymbol{\Sigma}$ is the
covariance matrix parametrized as
\begin{equation}
\boldsymbol{\Sigma} =
\left(\begin{array}{cc}
\sigma^2 (1+p) & c \sigma^2 \\
c \sigma^2 & \sigma^2(1-p)
\end{array}\right).
\end{equation}
Here $\sigma$ is the beamwidth, and $p$ and $c$ are the ellipticities
in the ``plus'' and ``cross'' directions respectively.  An elliptical
Gaussian with a major axis oriented along the $x'$ or $y'$ axes (see
Figure~\ref{fig:dparam_fpu}) has $+p$ or $-p$ ellipticity,
respectively, and one with a major axis oriented diagonally has $\pm
c$ ellipticity.  The total ellipticity is $e = \sqrt{p^2 + c^2}$.
Differential parameters (i.e. within a polarization pair) are defined
as differences of per-detector fits: differential beamwidth $d\sigma =
\sigma_A - \sigma_B$, differential pointing $dx = x_{0,A} - x_{0,B}$
\& $dy = y_{0,A} - y_{0,B}$, differential ellipticity $dp = p_A - p_B$
\& $dc = c_A - c_B$.  Here $A$ and $B$ refer to the orthogonally
polarized detectors within a pair.  Note that in this parametrization
differential beamwidth and differential pointing are expressed in
absolute units and not as fractions of the nominal beamwidth
\citep{wong14}.

Beam parameter statistics are generated by finding the best-fit
$\mathbf{x}$, $\sigma$, $p$, and $c$ for each beam in each mapping
run.  We then remove measurements in which the beam was not redirected
towards the source due to the mirror geometry (i.e. the receiver in
question was not physically underneath the mirror), the detector was
not operating normally, the fit did not converge for both $A$ and $B$,
the fit did not fall in a physically-acceptable range, or the residual
between the fit and the measured beam showed obvious artifacts.  In
a given year a typical \keckarray\ beam is measured $10$ times,
though this varies significantly across the focal plane.

For each detector and pair, we take the median across all measurements
as the best estimate of each parameter and take half the width of the
central 68\% of the distribution of those measurements as the
measurement uncertainty.\footnote{This statistic is relatively
  insensitive to outliers, and would equal $1\sigma$ for a Gaussian
  distribution of measurements.}  The characteristic uncertainty for
an individual measurement---taken to be the median of the measurement
uncertainties for all detectors/pairs across the array---is denoted
``Individual Measurement Uncertainty.''  In general the uncertainty on
per-detector ellipticity is somewhat large because random artifacts
occasionally escape the automated cuts and cause the fitting routine
to choose an ellipticity that is a poor fit to the beam.  Measurement
uncertainties of differential parameters are typically smaller than
those of per-detector parameters.  Common-mode effects, such as
artifacts in the timestreams or systematics associated with boresight
rotation, tend to affect both detectors in a pair equally and thus
bias parameter estimates in the same way.  We defer a detailed
discussion of noise and systematics in the beam measurement to
Section~\ref{ss:bmsyst}.

To characterize the distribution of parameters we also find the median
across the focal plane (``FPU Median'') and quantify the variation
across the focal plane as half the width of the central 68\% of the
distribution of best estimates for each detector/pair (``FPU
Scatter'').  Note that ``FPU Scatter'' measures the spread of
best-estimate parameters across the array and is not a measurement
uncertainty.  We correct for the non-negligible size of the thermal
source aperture when reporting $\sigma$
(Section~\ref{ss:beamprofile}).  

\subsection{Measured Beam Parameters}

\begin{table*}
\begin{center}
\caption{\keckarray\ 2014 Beam Parameter Summary Statistics}    
\begin{tabular}{|l|c|c|c|c|c|} 
\hline
\rule[-1ex]{0pt}{3.5ex}  &  Rx0 (95 GHz)  &  Rx1 (150 GHz) & Rx2 (95 GHz) &  Rx3 (150 GHz) & Rx4 (150 GHz) \\
\hline
$\sigma$ (${}^{\circ}$) & \ph{-}0.303 $\pm$ 0.003 $\pm$ 0.003      & \ph{-}0.213 $\pm$ 0.004 $\pm$ 0.003 & \ph{-}0.306 $\pm$ 0.002 $\pm$ 0.003 & \ph{-}0.216 $\pm$ 0.004 $\pm$ 0.003  & \ph{-}0.208 $\pm$ 0.002 $\pm$ 0.002 \\
$p$ (+)                & -0.010 $\pm$ 0.013 $\pm$ 0.019          & \ph{-}0.005 $\pm$ 0.021 $\pm$ 0.023 & -0.004 $\pm$ 0.011 $\pm$ 0.016 & \ph{-}0.008 $\pm$ 0.030 $\pm$ 0.021  & \ph{-}0.002 $\pm$ 0.018 $\pm$ 0.019\\
$c$ ($\times$)         & -0.001 $\pm$ 0.013 $\pm$ 0.020          & \ph{-}0.005 $\pm$ 0.022 $\pm$ 0.025 & -0.008 $\pm$ 0.011 $\pm$ 0.019 & -0.003 $\pm$ 0.012 $\pm$ 0.021 & \ph{-}0.005 $\pm$ 0.014 $\pm$ 0.023 \\
$dx$ ($'$)               & \ph{-0}0.46 $\pm$ \ph{0}0.76 $\pm$ \ph{0}0.05 & \ph{-0}0.10 $\pm$ \ph{0}0.97 $\pm$ \ph{0}0.06  & \ph{-0}0.34 $\pm$ \ph{0}0.64 $\pm$ \ph{0}0.06  & \ph{0}-0.11 $\pm$ \ph{0}0.35 $\pm$ \ph{0}0.08  & \ph{-0}0.18 $\pm$ \ph{0}0.42 $\pm$ \ph{0}0.05\\
$dy$ ($'$)               & \ph{0}-0.57 $\pm$ \ph{0}0.79 $\pm$ \ph{0}0.05  & \ph{0}-0.57 $\pm$ \ph{0}0.52 $\pm$ \ph{0}0.05 & \ph{0}-0.52 $\pm$ \ph{0}0.65 $\pm$ \ph{0}0.06 & \ph{0}-0.08 $\pm$ \ph{0}0.46 $\pm$ \ph{0}0.08    & \ph{0}-0.12 $\pm$ \ph{0}0.32 $\pm$ \ph{0}0.05\\
$d\sigma$ (${}^{\circ}$) & \ph{-}0.001 $\pm$ 0.002 $\pm$ 0.001    & \ph{-}0.001 $\pm$ 0.001 $\pm$ 0.001 & \ph{-}0.000 $\pm$ 0.001 $\pm$ 0.001 &  \ph{-}0.000 $\pm$ 0.001 $\pm$ 0.001 & \ph{-}0.000 $\pm$ 0.001 $\pm$ 0.001 \\
$dp$ (+)               & -0.013 $\pm$ 0.014 $\pm$ 0.002         & -0.009 $\pm$ 0.010 $\pm$ 0.002 & -0.006 $\pm$ 0.004 $\pm$ 0.002 &  -0.007 $\pm$ 0.012 $\pm$ 0.003 & -0.020 $\pm$ 0.004 $\pm$ 0.002\\
$dc$ ($\times$)        & \ph{-}0.002 $\pm$  0.004 $\pm$ 0.002   & -0.010 $\pm$ 0.005 $\pm$ 0.003 & \ph{-}0.002 $\pm$ 0.003 $\pm$ 0.002 & -0.002 $\pm$ 0.008 $\pm$ 0.002 & -0.002 $\pm$ 0.002 $\pm$ 0.002\\
\hline
\end{tabular}
\tablecomments{Beam parameters are listed as FPU median $\pm$ FPU
  scatter $\pm$ individual measurement uncertainty.  See
  Section~\ref{sec:gauss} for a detailed discussion of the
  uncertainties.  \cite{biceptwoIV} presents analogous statistics for
  \biceptwo\ and \keckarray\ in previous seasons.}
\label{tab:k14_beamstats}
\end{center}
\end{table*}

\begin{table*}
\begin{center}
\caption{\keckarray\ 2015 Beam Parameter Summary Statistics}    
\begin{tabular}{|l|c|c|c|c|c|} 
\hline
\rule[-1ex]{0pt}{3.5ex}  & Rx0 (95 GHz) & Rx1 (220 GHz) & Rx2 (95 GHz)& Rx3 (220 GHz) & Rx4 (150 GHz) \\
\hline
$\sigma$ (${}^{\circ}$)   & \ph{-}0.304 $\pm$ 0.003 $\pm$ 0.003  & \ph{-}0.141 $\pm$ 0.002 $\pm$ 0.002 & \ph{-}0.307 $\pm$ 0.002 $\pm$ 0.004  & \ph{-}0.142 $\pm$ 0.002 $\pm$ 0.003  & \ph{-}0.207 $\pm$ 0.003 $\pm$ 0.002\\
$p$ (+)                 & -0.007 $\pm$ 0.015 $\pm$ 0.019 &       \ph{-}0.003 $\pm$ 0.020 $\pm$ 0.028 & -0.003 $\pm$ 0.011 $\pm$ 0.017 & \ph{-}0.001 $\pm$ 0.022 $\pm$ 0.031  & \ph{-}0.002 $\pm$ 0.022 $\pm$ 0.017\\
$c$ ($\times$)          & -0.002 $\pm$ 0.013 $\pm$ 0.019 &       -0.003 $\pm$ 0.021 $\pm$ 0.031      & -0.005 $\pm$ 0.012 $\pm$ 0.019  & \ph{-}0.002 $\pm$ 0.028 $\pm$ 0.035  & \ph{-}0.005 $\pm$ 0.017 $\pm$ 0.016\\
$dx$ ($'$)                & \ph{-0}0.42 $\pm$ \ph{0}0.76 $\pm$ \ph{0}0.04  & \ph{-0}0.32 $\pm$ \ph{0}0.26 $\pm$ \ph{0}0.04 & \ph{-0}0.33 $\pm$ \ph{0}0.68 $\pm$ \ph{0}0.05  & \ph{0}-0.49 $\pm$ \ph{0}0.19 $\pm$ \ph{0}0.04  & \ph{-0}0.19 $\pm$ \ph{0}0.40 $\pm$ \ph{0}0.04\\
$dy$ ($'$)                & \ph{0}-0.56 $\pm$ \ph{0}0.78 $\pm$ \ph{0}0.04  & \ph{0}-0.51 $\pm$ \ph{0}0.23 $\pm$ \ph{0}0.04 & \ph{0}-0.54 $\pm$ \ph{0}0.65 $\pm$ \ph{0}0.05  & \ph{-0}0.43 $\pm$ \ph{0}0.12 $\pm$ \ph{0}0.03  & \ph{0}-0.11 $\pm$ \ph{0}0.33 $\pm$ \ph{0}0.03\\
$d\sigma$ (${}^{\circ}$) & \ph{-}0.001 $\pm$ 0.002 $\pm$ 0.001  & \ph{-}0.000 $\pm$ 0.000 $\pm$ 0.001 & \ph{-}0.000 $\pm$ 0.001 $\pm$ 0.001 & \ph{-}0.000 $\pm$ 0.000 $\pm$ 0.001 & \ph{-}0.000 $\pm$ 0.001 $\pm$ 0.001 \\
$dp$ (+)               & -0.013 $\pm$ 0.013 $\pm$ 0.002        & -0.015 $\pm$ 0.009 $\pm$ 0.006     & -0.006 $\pm$ 0.004 $\pm$ 0.002 &  -0.017$\pm$ 0.005 $\pm$ 0.005 & -0.019 $\pm$ 0.005 $\pm$ 0.002\\
$dc$ ($\times$)         & \ph{-}0.002 $\pm$ 0.004 $\pm$ 0.002  & \ph{-}0.001 $\pm$ 0.006 $\pm$ 0.006 & \ph{-}0.002 $\pm$ 0.002 $\pm$ 0.002 & \ph{-}0.004 $\pm$ 0.005 $\pm$ 0.005  & -0.002 $\pm$ 0.002 $\pm$ 0.002 \\
\hline
\end{tabular}
\tablecomments{Beam parameter are listed as FPU median $\pm$ FPU
  scatter $\pm$ individual measurement uncertainty.  See
  Section~\ref{sec:gauss} for a detailed discussion of the
  uncertainties.  Only Rx0/Rx2/Rx4 are in common with
  Table~\ref{tab:k14_beamstats}.  \cite{biceptwoIV} presents analogous statistics for
  \biceptwo\ and \keckarray\ in previous seasons.}
\label{tab:k15_beamstats}
\end{center}
\end{table*}

Tables~\ref{tab:k14_beamstats}
and~\ref{tab:k15_beamstats} show per-detector and differential beam
parameters for \keckarray\ receivers in 2014 and 2015, respectively,
in the following format: FPU Median $\pm$ FPU Scatter $\pm$ Individual
Measurement Uncertainty.

Beamwidths for the three frequencies in the BK15 dataset are roughly
$0.305^{\circ}$ ($43'$ FWHM, 95\,GHz), $0.210^{\circ}$ ($30'$ FWHM,
150\,GHz), and $0.141^{\circ}$ ($20'$ FWHM, 220\,GHz).  Variability
across the focal plane is not detected above the individual
measurement uncertainty.  The central values of per-detector $p$ and
$c$ ellipticities are generally close to zero and are spread equally
above and below; this is because the optical design places the optimal
focus on an annulus of detectors located a median distance from the
center of the focal plane, causing enhanced ellipticity towards the
edge.  Such a pattern is roughly azimuthally symmetric, so that $p$
and $c$ average to small values.

\emph{Differences} in beam shapes between the orthogonally polarized
$A$ and $B$ detectors contribute to $T \rightarrow P$ leakage
\citep{hu03}.  The majority of the power in the
\bicep/\keck\ difference beams is encapsulated in a second-order
expansion of the beam profile, which couples to the CMB temperature
sky and its first and second derivatives.  The modes corresponding to
these couplings are either ``deprojected'' from the CMB maps by
scaling and removing the best-fit templates of the \planck\ sky map
and its derivatives from pair difference data, or ``subtracted'' by
removing these templates with amplitudes determined directly from beam
maps.  We typically subtract differential ellipticity instead of
deprojecting it, because deprojecting would preferentially filter the
$TE$ and $EE$ spectra.  \bicep/\keck\ polarization maps are therefore
largely insensitive to $T \rightarrow P$ leakage entering through
these modes; see BK-III for more details.\footnote{This assumes the
  \planck\ temperature maps are free of systematics.  We have also
  simulated the effect of \planck\ noise in the templates and found it
  to be negligible (BK-III).  Noise in beam measurements could affect
  the accuracy of subtraction; based on the repeatibility of $dp$ and
  $dc$, the uncertainty in differential ellipticity could contribute
  $T \rightarrow P$ leakage at the $\rho < 1\times 10^{-6}$ level (see
Section~\ref{ss:bmsim}).}

Nevertheless, since differential Gaussian parameters---which
correspond roughly to the modes that are
deprojected\footnote{Deprojection coefficients are obtained by
  regressing the pair difference data against templates of the
  \planck\ temperature sky and its derivatives that have been smoothed
  by our array-averaged beam profile.  If the beam profile were
  perfectly Gaussian the coefficients would exactly match the
  parameters obtained by differencing Gaussian fits.}---often probe
optical and detector fabrication effects, it is worthwhile to quantify
and plot them in focal plane coordinates to understand their causes
and how they propagate through the final analysis.  Here we discuss
differential pointing and ellipticity, which if left uncorrected could
contribute significantly to $T \rightarrow P$ leakage.  Differential
beamwidth is generally small enough to be negligible.

\begin{figure*}[h]
\begin{center}
\resizebox{0.95\textwidth}{!}{\includegraphics{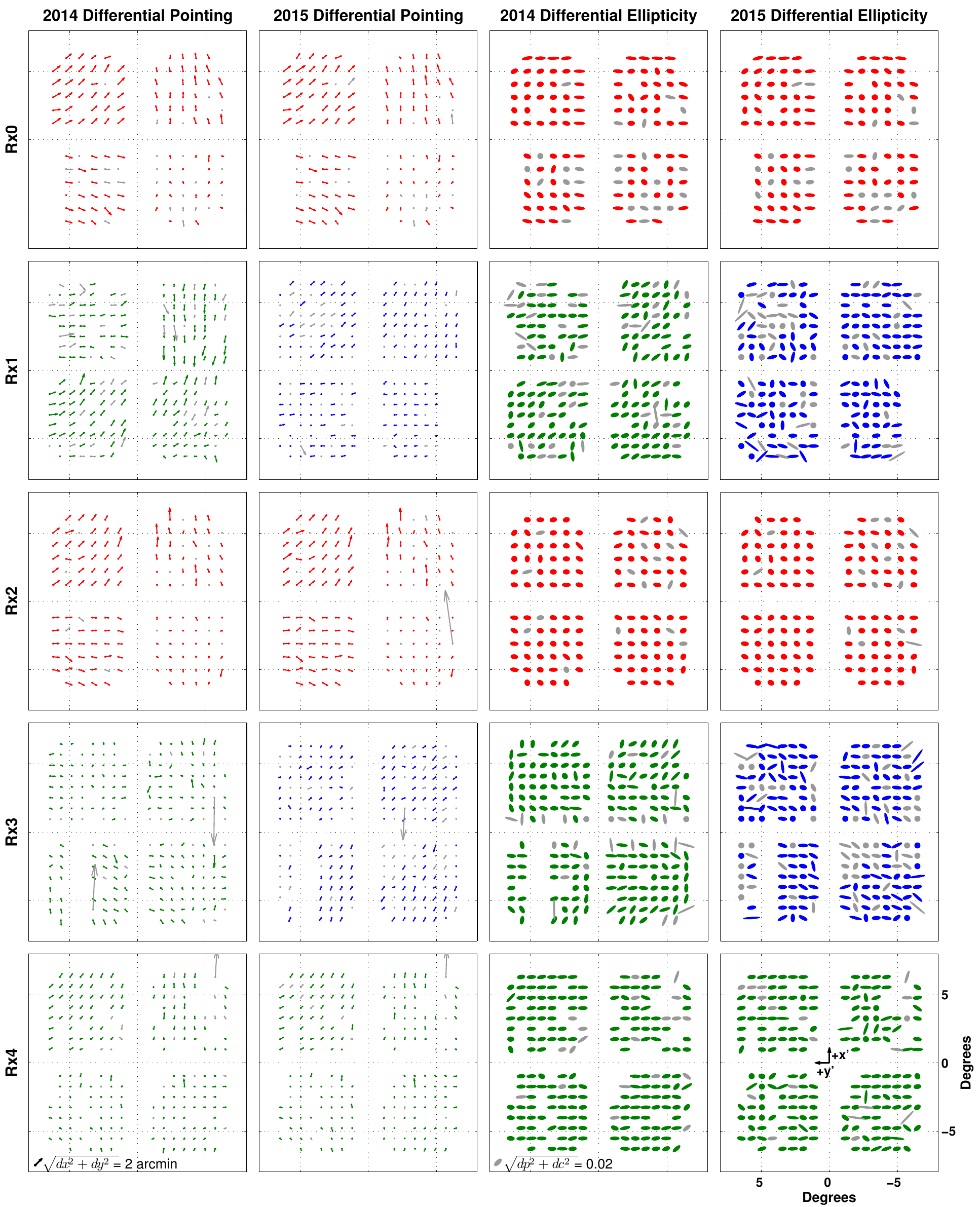}}
\end{center}
\caption{Differential pointing and ellipticity for
  \keckarray\ receivers in the 2014 and 2015 seasons.  Parameters are
  plotted in focal plane coordinates as projected on the sky.  $A$/$B$
  detectors are polarized along the $y'$/$x'$ axes respectively,
  indicated in the bottom right panel.  Detectors at 95\,GHz are in
  red, 150\,GHz in green, and 220\,GHz in blue; those in gray are not
  used in analysis.  Parameters are consistent across years when the
  receiver did not change---Rx0, Rx2, and Rx4 stayed the same in 2014
  and 2015.  \emph{Left columns}: Differential pointing rendered as an
  arrow pointing from the $A$ detector location to the $B$ detector
  location.  The arrow length indicates the degree of mismatch,
  multiplied by 20.  \emph{Right columns}: Differential ellipticity
  rendered as ellipses; major axes are proportional to $\sqrt{dp^2 +
    dc^2}$, a measure of the magnitude of the differential ellipticity
  (minor axes are fixed).  The magnitude has been multiplied by 75 for
  visibility.  \cite{biceptwoIV} presents analogous plots for
  \biceptwo\ and \keckarray\ in previous seasons.}
\label{fig:dparam_fpu}
\end{figure*}

Figure~\ref{fig:dparam_fpu} shows differential pointing $dx,dy$ and
ellipticity $dp,dc$ for all \keckarray\ 2014 and 2015 receivers in
focal plane layout.  Differential pointing is usually the beam shape
mismatch mode that causes the most $T \rightarrow P$
leakage.\footnote{While relative gain is also important, it is not
  measured in beam mapping because beam map normalization effectively
  deprojects it.  For CMB analysis, relative gain is corrected at the
  timestream level by normalizing by the response to a small change in
  atmospheric loading (``elevation nods''), and at the map level by
  deprojection.}  Several trends stand out: first, measurements are
consistent between years in which the receiver did not change
(e.g. Rx0 in 2014 and 2015).  Second, the absolute magnitude is
somewhat frequency-dependent---95\,GHz receivers show larger offsets
than both 150 and 220\,GHz, which are similar to each other---but
there is large variability.  Finally, it is typical to have a coherent
differential pointing direction across a detector tile.  Significant
effort at the design and fabrication stages has been spent in reducing
this mode, resulting in tiles with factors of $\sim$\,2--4 lower
differential pointing than earlier, \biceptwo-era tiles
\citep{obrient12}.  Differential ellipticity is similarly consistent
from year to year and coherent within tiles; $-dp$ appears to
dominate.

We have checked that the deprojection coefficients---which scale
templates of the \planck\ temperature sky and its derivatives to best
fit the CMB data---match measured beam parameters and exhibit scatter
consistent with their noise.  On-sky data confirm that the CMB maps
and far-field beam maps measure the same dominant beam mismatch modes
(BK-III).

It is the higher-order residuals, not captured by the above
second-order expansion of the beam profile, which could contribute $T
\rightarrow P$ leakage to the final polarization maps.  To move beyond
differential Gaussians we require sensitive beam maps capable of
capturing the potentially-complex higher-order difference beam
morphology.

\section{Composite Beam Maps and Beam Profiles}
\label{sec:composites}

To fully assess the impact of beams in CMB analysis we use deep
far-field measurements extending to several degrees away from the beam
center.  In this section we describe how the beam maps generated in
Section~\ref{sec:ffbm}---hereafter denoted ``component'' maps---are
combined to form high-fidelity ``composite'' maps.  Using composite
beam maps reduces noise and systematics in the beam measurement and
allows us to measure the main beam at all azimuthal angles.

\subsection{Noise and Systematics in Beam Measurements}
\label{ss:bmsyst}

Typical noise levels in a single pair's component beam maps (i.e. from
one measurement made with the 45\,cm chopper) would induce a bias on
the prediction of $T \rightarrow P$ leakage from mismatched beams
equivalent to $\rho \sim 0.02 \pm 0.01 $ at 150\,GHz even if the beams
were perfectly matched\footnote{Beam map noise is measured by taking
  the standard deviation of pixel values in a region of the map far
  away from the main beam and ground contamination (i.e. more than
  $10^{\circ}$ above the beam).  For 150\,GHz detectors measured with
  the 45\,cm chopper, the typical rms noise level in $0.1^{\circ}$
  pixels is 1/800 of the main beam amplitude.  The bias on $r$ is
  estimated by generating Gaussian noise simulations with these
  amplitudes and running them through the beam map simulation
  pipeline.} (see Section~\ref{ss:bmsim}).  This bias and uncertainty
can be reduced by combining maps from multiple measurements.  The
average pair has $\sim 10$ good maps per year
(Section~\ref{sec:gauss}), so assuming Gaussian noise and a systematic
error-free measurement, we expect to measure leakage with an
uncertainty of $\sigma(r) \sim 0.001$ for a \emph{single} pair in a
given year after averaging.  Similarly---since our estimate of $T
\rightarrow P$ leakage for the CMB map is made by coadding the
predicted $Q/U$ leakage maps from all pairs---if noise is uncorrelated
across detectors, the uncertainty on the final leaked $BB$ estimate
scales down with the number of pairs.  Therefore, if there is real
structure in the higher-order difference beams at a level that could
be important for CMB analysis, we fully expect the beam maps to
measure it.  Since the 45\,cm chopper noise level already outperforms
our $\sigma(r)$ requirements, the current 60\,cm thermal chopper will
be well below the requirement for the next generation of
\bicep\ experiments.

The most prominent systematic in the beam map measurement is
ground-fixed contamination, which leaks into the in-phase demodulated
signal at a low level.\footnote{This contamination arises when
  scanning across a change in temperature, such as from cold sky to
  the (relatively) warm South Pole Telescope.  Part of that
  temperature change is interpreted by the deconvolution kernel as
  signal from the thermal chopper.  Ground-fixed signal usually enters
  into the measurement at the $-20$ dB level or lower.}  We
conservatively choose to mask out all regions of the map with known
ground-fixed structure.  To measure the regions of the beam that were
masked, the receiver is rotated about the boresight axis and another
map is made.

Other potential systematics include curvature in the redirecting
mirror, the finite distance of the source, and nonlinearity in the
detector response at the relatively high loading conditions used in
calibration measurements.  Using the measured mirror
curvature\footnote{Using \emph{in situ} photogrammetry we have
  constrained the mirror's flatness to better than 1.5 mm across its
  surface.} we find that the absolute beam shape experiences a very
small change---$B(\ell)$ changes by $< 0.01 \%$ for the 220\,GHz beams
at $\ell = 200$---and the differential beam shapes will see even less
distortion because both detectors in a pair reflect off the mirror
nearly identically.  Through simulation of the beam shape error
expected for a measurement at 210\,m instead of at infinity, we find
that BICEP3 with a nominal far-field distance of 171\,m experiences a
deviation of $<0.2\%$ at $\ell = 200$; the effect will be smaller for
all \keckarray\ bands.  Finally, we constrain nonlinearity on the
high-loading aluminum transition sensor to be $\lesssim 1\%$ using
load curves in which the detector bias voltage is ramped, providing an
I-V characteristic.

Nonrepeatable contamination, i.e. non-Gaussian noise, is more
worrisome and most often consists of single-sample glitches, which
manifest as ``hot spots'' in the demodulated timestreams.  A single
glitch in the timestream can have an amplitude much larger than that
of the main beam, resulting in a leakage prediction orders of
magnitude larger than the true value even if the rest of the beam map
were accurate.

For this reason we regard the differential beam maps as upper limits
on the $T \rightarrow P$ leakage from mismatched beams: they should
capture all true leakage within the beam map radius, but could also
contain non-Gaussian noise or measurement systematics that are
difficult to quantify.  Looking towards future results, we have
demonstrated a reduction in beam map contamination through improved
low-level deglitching and data quality cuts.  These improvements will
find application both in reanalysis of existing beam map data and in
future datasets.

\subsection{Composite Beam Map Generation}
\label{ss:composite_gen}

\begin{figure*}
\begin{center}
\resizebox{0.8\textwidth}{!}{\includegraphics{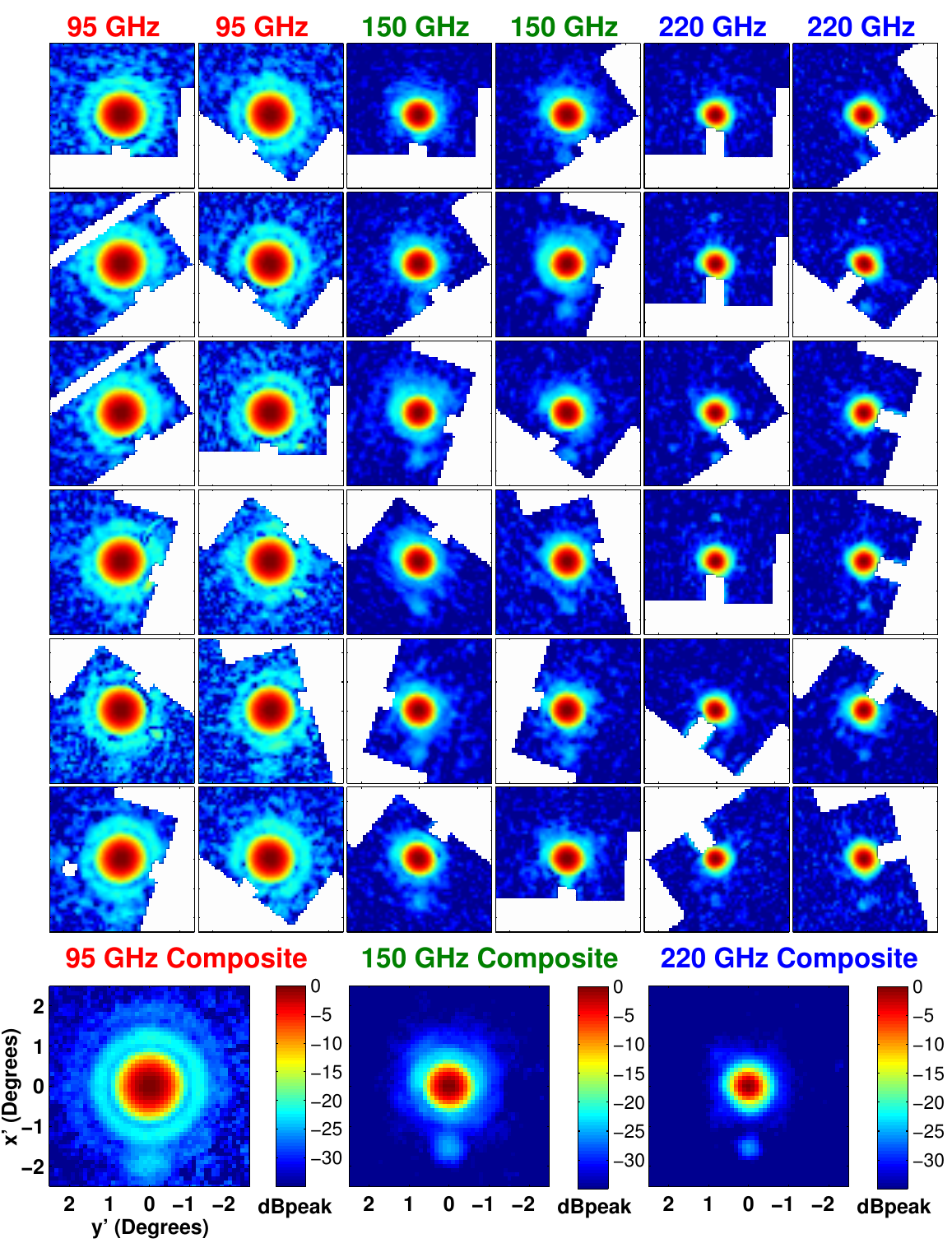}}
\end{center}
\caption{\emph{Large panels:} Example composite beam maps for
  individual 95\,GHz (left), 150\,GHz (middle), and 220\,GHz (right)
  detectors.  \emph{Small panels:} the component maps that contribute
  to the composite, taken at multiple boresight angles---the ground,
  mast, and SPT have been masked out.  The $x',y'$ coordinate system
  is detector-fixed.  The circular feature at the $\sim -23$ dB level,
  which in all cases here appears below the main beam, is due to
  crosstalk in the readout system and has been extensively
  characterized \citep{biceptwoII, biceptwoIII}.  The maps are in dB
  relative to peak amplitude and share the same color scale; because
  beams of different widths have different peak amplitudes, the noise
  in the lower-frequency beams appears inflated.}
\label{fig:composites}
\end{figure*}

Composite beam maps are generated by combining component maps that
have passed data quality cuts and in which known, spatially-fixed
contamination has been removed.  We begin with the set of component
maps from which beam statistics were calculated
(Section~\ref{sec:gauss}).  We then apply a spatial mask: the ground
($>2^{\circ}$ below the source), the mast, and the South Pole
Telescope are removed.  Maps are centered on the common pair centroid,
rotated to account for boresight angle, and peak-normalized.  Finally
the composite is made by assembling a stack of all good maps and
taking a median in each spatial pixel.  We take a median instead of a
mean because occasionally high-amplitude artifacts escape the
automated cuts.  In current \keckarray\ maps, median-filtering results
in demonstrably lower noise levels and fewer artifacts than
mean-filtering.  In future datasets we plan to improve low-level
deglitching and spatial coverage to the point at which a mean may be
taken.

Figure~\ref{fig:composites} shows sample composite beam maps and the
component maps used to form them at the three \keckarray\ frequencies.
The component maps have been masked for ground-fixed signal.  Noise
reduction and rejection of spurious contaminating signals are evident
in the composites.

In principle it is best to use measurements that extend far away from
the beam center to capture as much of the optical response as
possible.  We note, however, that in standard CMB observations the
comoving forebaffle absorbs off-axis response at large angles
(generally $> 10^{\circ}$ from the main beam).  Since the forebaffles
are removed to make room for hoisting the far-field flat mirror,
response measured in the far-field beam maps at very large angles may
not correspond to real beam pickup during CMB scans.\footnote{In
  separate measurements, we use an amplified microwave source to
  measure far sidelobes.  Comparing such maps made with and without
  the forebaffle, we see no perceptible difference in the main beam
  region.  We expect any such difference to be extremely small, given
  the small amount of power intercepted ($\sim 0.7\%$, see BK-IV) and
  the very efficient absorption of this power by the forebaffle.}  The
composite maps also become noisier further away from the beam center
because there are fewer hits per pixel as a result of the spatial
masking.  For the beam map simulation we use composites made out to
$2^{\circ}$ from the beam center (Section~\ref{sec:ttop}).

\subsection{Beam Profiles}
\label{ss:beamprofile}

\begin{figure*}
\begin{center}
\resizebox{0.9\textwidth}{!}{\includegraphics{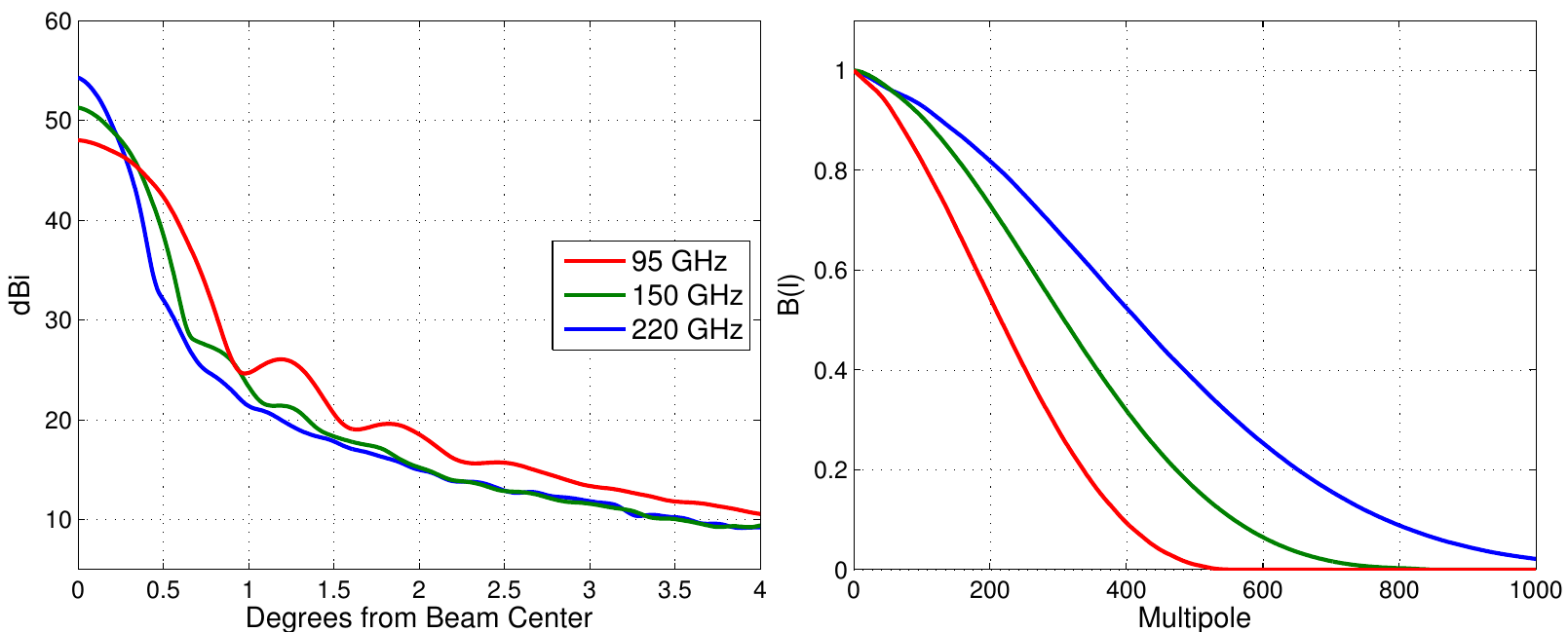}}
\end{center}
\caption{\emph{Left}: Azimuthally averaged beam profiles for
  \keckarray\ at 95, 150, and 220\,GHz, coadded over all operational
  channels in 2015.  The maps are normalized relative to an isotropic
  radiator under the assumption that all beam power is contained
  within the $r < 4^{\circ}$ maps.  \emph{Right}: The $B(\ell)$ window
  functions, equivalent to Fourier transforms of the beam profiles.
  The effect of the finite source size has been removed.
  \cite{biceptwoIV} presents analogous plots for \biceptwo\ and
  \keckarray\ in previous seasons.  }
\label{fig:profilebl}
\end{figure*}

To generate array-averaged beam profiles we coadd the composite maps
over all detectors contributing to CMB data, inversely weighted by
CMB-derived per-detector noise.  Figure~\ref{fig:profilebl} shows the
averaged radial profiles and the equivalent $B(\ell)$ window
functions.  To remove the effect of the finite size of the source
aperture, we divide $B(\ell)$ by $\frac{ 2J_1 (\ell D/2)}{\ell D/2}$,
where $J_1$ is the Bessel function of the first kind and $D$ is the
angular diameter of the source as seen from the telescope.  Typical
detector-to-detector variations at $\ell=100$ are 2.6\% (95\,GHz),
2.2\% (150\,GHz), and 3.7\% (220\,GHz); the statistical uncertainty on
individual beam profiles is small compared to this variation.  We use
these beam profiles in CMB analysis to smooth input maps for signal
simulations and to smooth the \planck\ temperature sky and its
derivatives, which serve as deprojection templates.

\
\

\section{Simulation of Temperature-to-Polarization Leakage}
\label{sec:ttop}

To estimate the $T\rightarrow P$ leakage present in the BK15 maps we
run ``beam map simulations'' with composite beam maps such as those
presented in Section~\ref{sec:composites}.\footnote{Although the term
  ``simulation'' is used, all inputs are based on real data:
  \planck\ maps, detector pointing, and on-sky beams measured \emph{in
    situ} at South Pole.}  In this section we review how the
simulations are generated and show the predicted leaked polarization
maps and power spectra.  The auto power spectra represent upper limits
to the $T \rightarrow P$ leakage in the BK15 maps due to mismatched
beam shapes within polarization pairs.  We then take the cross spectra
of the beam map simulations with the real BK15 maps, which represent
our best estimate of the leakage present in CMB data.

\subsection{Simulation Methodology and Leakage Estimates}
\label{sec:bmsim}

The standard BK15 pipeline is used to run beam map simulations.  We
inject $T\rightarrow P$ leakage at the timestream level by convolving
the \planck\ temperature sky with the per-detector beam maps and
sampling from the resulting smoothed maps using real detector pointing
data.\footnote{The detector-centered, locally-flat beam maps are
  scaled and rotated using the full spherical geometry appropriate for
  each detector's pointing trajectory prior to convolution with a
  flat-sky projection of the \planck\ map.  The \planck\ beam has been
  deconvolved from the temperature map.  } We intentionally set the
\planck\ $Q/U$ maps to zero so that any measured polarization is a
result of $T \rightarrow P$ leakage from mismatched beams.  The
timestreams are then processed into $Q/U$ maps just as are the real
data, including identical cuts and detector weighting.  Deprojection
of differential gain/pointing and subtraction of differential
ellipticity are also applied\footnote{There is a small mismatch
  between the simulation timestreams (formed from flat-sky
  convolutions) and the deprojection templates (formed from curved
  \healpix\ maps), but we have demonstrated that it corresponds to
  $\rho \sim 10^{-6}$ and is negligible; see BK-III for details.} (see
Section~\ref{sec:gauss}).  As in our standard \bmode\ analysis,
bandpowers are measured after applying a matrix-based purification
that reduces $E \rightarrow B$ leakage from partial sky coverage and
filtering \citep{biceptwoVII} so that only the leakage modes relevant
for our actual CMB maps are used.  All beams contributing to the BK15
maps, including \biceptwo\ and \keck\ 2012--2013 \citep{biceptwoIV},
are accounted for in these results.

We assess the impact of a systematic contribution to a
single-frequency $BB$ spectrum with a quadratic estimator $\rho$ that
is the equivalent $r$ level of the contamination.  The estimator is
\begin{equation}
\rho = \frac{\langle \mathbf{C} \rangle^T \mathbf{N}^{-1} \hat{\mathbf{C}}}{ \langle \mathbf{C} \rangle^T \mathbf{N}^{-1} \langle \mathbf{C} \rangle },
\end{equation}
where $\hat{\mathbf{C}}$ are the systematic bandpowers predicted from
the beam map simulation, $\mathbf{N}$ is the bandpower covariance
matrix from signal + noise simulations (``signal'' refers to our
fiducial lensed-\lcdm\ + dust skies and ``noise'' matches that of the
BK15 maps), and $\langle \mathbf{C} \rangle$ are the $BB$ bandpower
expectation values for an $r=1$ signal.  This effectively weights the
systematic bandpowers by the ratio of the expected signal to the noise
variance in each $\ell$ bin.  For \bicep/\keckarray, most of the
statistical power in the $\rho$ estimate comes from the first three
bins \citep{barkats14}.

\subsection{Simulation Results}
\label{ss:bmsim}

\begin{figure}
\resizebox{\columnwidth}{!}{\includegraphics{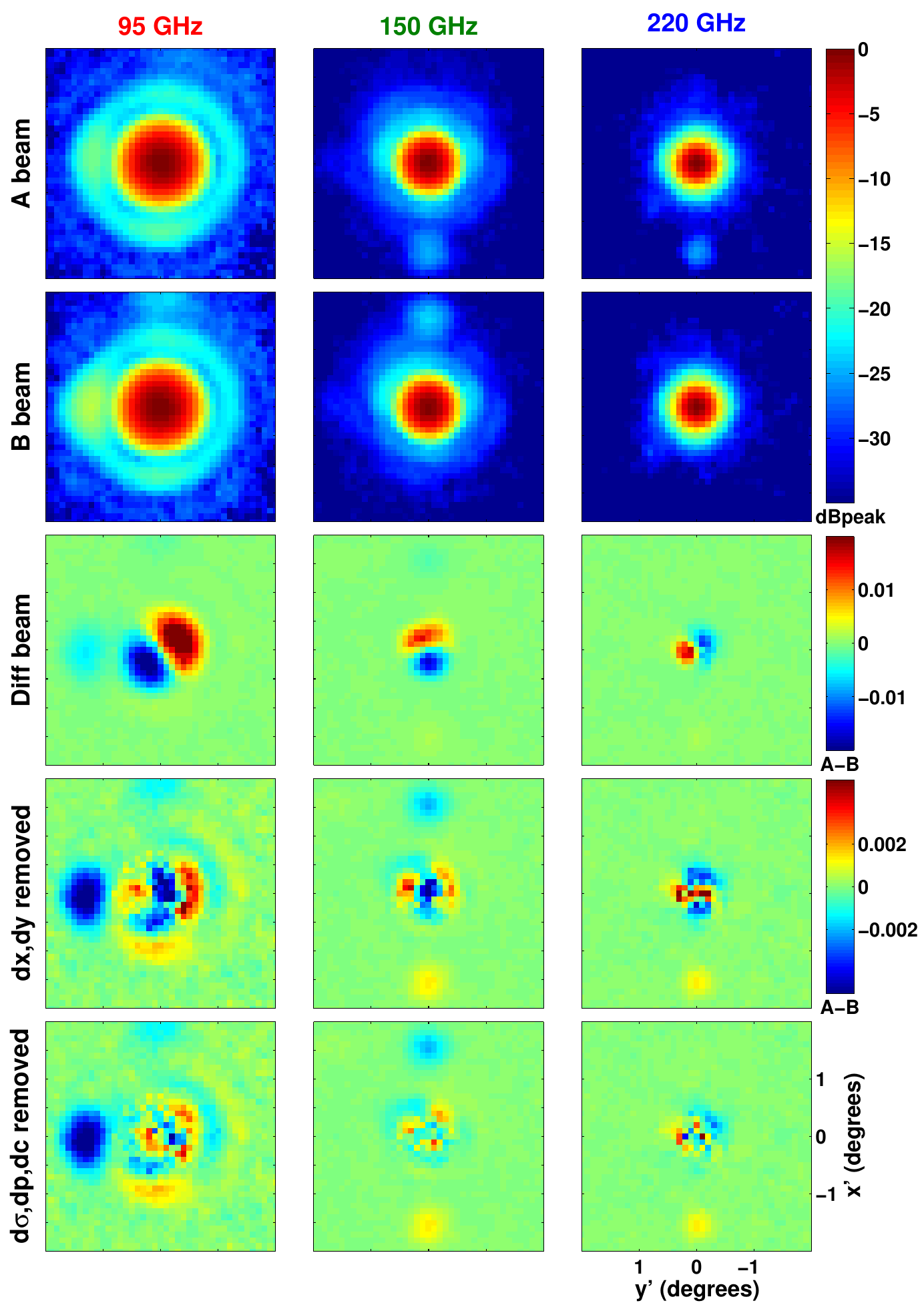}}
\caption{Example differential beam maps at 95\,GHz (left), 150\,GHz
  (middle), and 220\,GHz (right).  In each column, the top two panels
  show composite maps of $A$ and $B$ detectors in a pair; the color
  scale is in dB relative to the peak amplitude.  The third panel
  shows the difference between $A$ and $B$, dominated by differential
  pointing.  The color scale is linear relative to the pair sum peak
  amplitude ($\pm 2\%$)---this has effectively had differential gain
  deprojected.  The fourth panel shows the residual after differential
  pointing is deprojected; note the color scale has changed ($\pm
  0.5\%$).  The last panel has the same color scale as the fourth and
  shows the undeprojected residual after differential pointing,
  beamwidth, and ellipticity have been removed; this contributes to
  the $T \rightarrow P$ leakage discussed in Section~\ref{ss:bmsim}.
  The large feature in the 95\,GHz map is discussed in more detail in
  the text.}
\label{fig:diffbeams}
\end{figure}

\begin{figure*}
\begin{center}
\resizebox{0.95\textwidth}{!}{\includegraphics{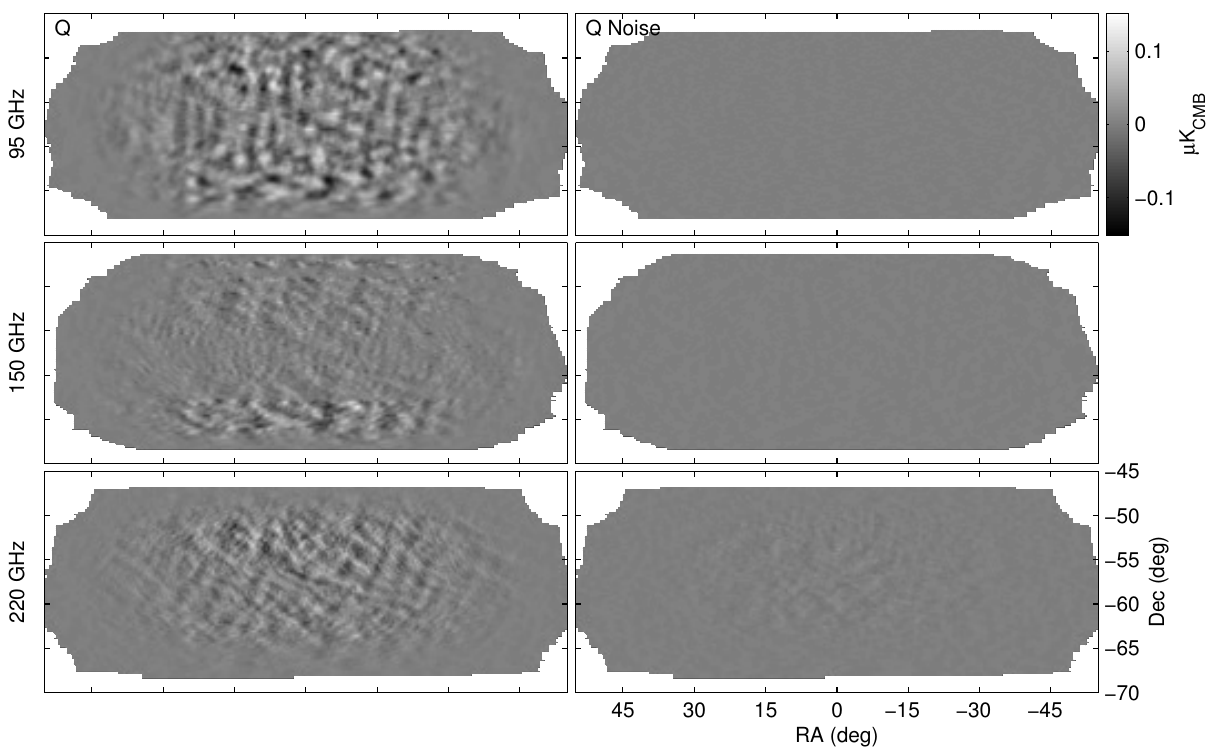}}
\end{center}
\caption{Apodized $Q$ maps of $T \rightarrow P$ leakage predicted by
  the beam map simulations (left column) and a beam map noise
  realization (right column).  The visible signal is due to
  higher-order undeprojected residuals in the measured differential
  beam maps and potentially systematics in the beam measurement.  The
  leakage at 95\,GHz is mostly $E$ mode in character and is due to the
  anomalous feature in the 95\,GHz tile edge pixels, shown in
  Figure~\ref{fig:diffbeams}.  The band of lower leakage at the center
  of the 150\,GHz map is due to the complex cancellation of the 17
  receiver-years that compose the map.}
\label{fig:bmsim_qu}
\end{figure*}

Only the component of the pair difference beam remaining after
deprojection (the ``undeprojected residual'') is relevant for
assessing the main beam $T \rightarrow P$ leakage in the BK15 map.  To
illustrate typical difference beams, in Figure~\ref{fig:diffbeams} we
plot example composite maps for both $A$ and $B$ detectors in a pair,
the pre-deprojection difference map, the difference map with
differential pointing removed, and the final undeprojected
residual.  Amplitudes of features in typical undeprojected residual
maps are $\sim 0.2\%$ of the main beam peak.

Far-field beam maps routinely identify features in individual detector
pairs that contribute excess $T \rightarrow P$ leakage.  To
illustrate, we show a 95\,GHz beam that contains a large feature in
the undeprojected residual, caused by a well-characterized anomalous
interaction between the detector tile corrugations and the tile edge
detectors.  The impact on the undeprojected residuals and the
subsequent remedy is discussed in more detail in
Section~\ref{ss:bmsim_discuss}.

The $Q$ maps of residual $T \rightarrow P$ leakage predicted using the
composite beam maps are shown in Figure~\ref{fig:bmsim_qu}.  An
estimate of the beam map noise is also shown, and discussed in more
detail below.  We expect all of the true leakage to be captured in
these maps, but again emphasize that there is likely a nonnegligible
contribution from systematics in the beam measurement.

While in general there is no expectation that undeprojected residuals
should prefer $E$ or $B$, it is not uncommon for the leakage to
contaminate one over the other. For example, the 95\,GHz tile
corrugation feature shown in Figure~\ref{fig:diffbeams} leaks
primarily to $E$ modes because it is aligned with the detector
polarization axes (the $Q$ beam)---leakage aligned at $45^{\circ}$
would leak to $B$ (see BK-III for more details).

The central region of the 150\,GHz map contains a band of lower
leakage, which is due to the complex interaction of the rotational
symmetry of undeprojected residuals (which determines whether they
cancel under boresight rotation), their distribution across the focal
plane (which determines whether they cancel when coadded with other
detectors), and the combination of the many distinct receivers that
contribute to the map.

\begin{figure*}
\begin{center}
\resizebox{0.95\textwidth}{!}{\includegraphics{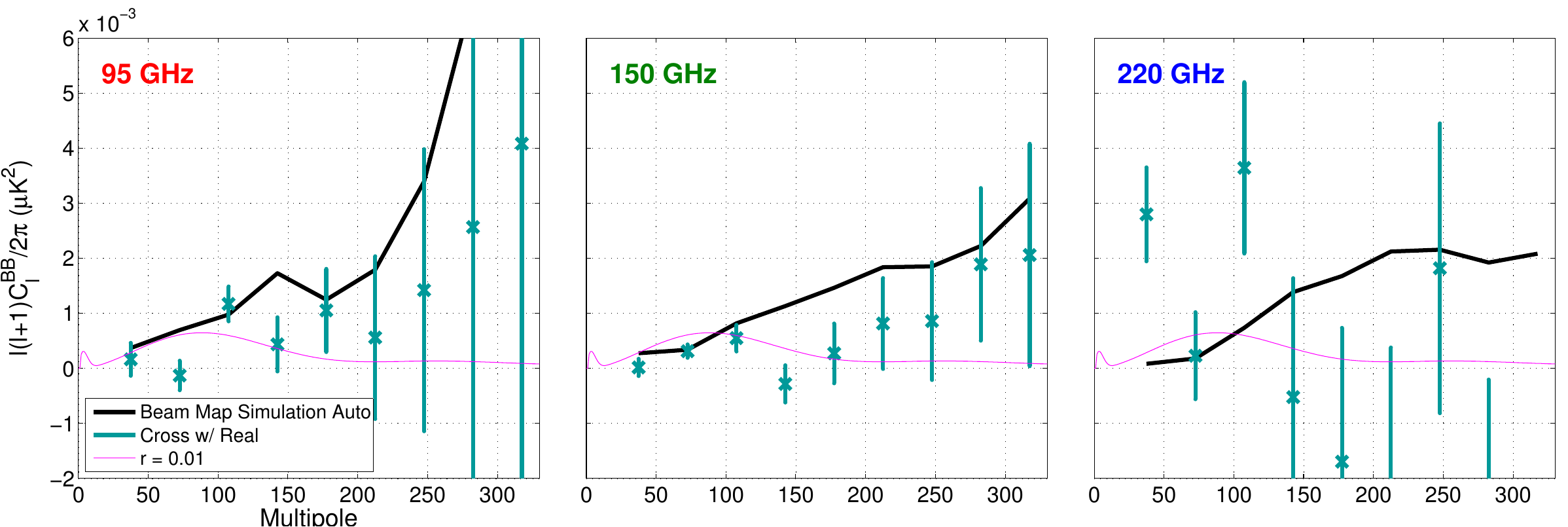}}
\end{center}
\caption{$BB$ power spectra from beam map simulations, corresponding
  to the maps shown in Figure~\ref{fig:bmsim_qu}.  The black lines are
  the per-frequency auto spectra, which have been noise-debiased using
  the beam map jackknife maps.  They should be considered upper limits
  on $T \rightarrow P$ leakage.  The teal crosses show the cross
  spectra of the beam map simulations with the BK15 maps.  The error
  bars are derived from the cross spectra of the fixed beam map
  simulation with 499 CMB lensing + dust + instrumental/atmospheric
  noise simulations.  Noise and systematics in the beam map
  measurement are not included in this error estimate.}
\label{fig:bmsim_result}
\end{figure*}

The BB spectra corresponding to these maps are shown in
Figure~\ref{fig:bmsim_result}.  We present both noise-debiased auto
spectra and cross spectra of the beam map simulations with the BK15
maps. The interfrequency cross spectra, corresponding to leakage that
correlates across frequencies, are consistent with zero within the
uncertainties and are not plotted here.

The beam map auto spectra (black lines) represent upper limits to the
single-frequency leakage since the beam measurement may contain
low-level systematics.  We estimate the noise contribution by forming
a ``beam map jacknife'' for each detector pair.  For each beam, we
randomly divide the component maps into two halves and form two
separate composite maps.  The difference between these maps is
effectively a jackknife in which the signal is removed and the noise
remains (Figure~\ref{fig:bmsim_qu} right column).  This measure of
noise bias has been subtracted from the beam map auto spectrum.  The
$\rho$ estimates for the upper limits and the jackknives are shown in
Table~\ref{tab:bmsim_rho} and indicate that the noise contribution to
the beam maps is subdominant compared to the combination of true beam
mismatch and potential systematics in the measurement.

The cross spectra with the BK15 maps (teal crosses; see BK-X Figures
7--9) offer an unbiased estimate of $T \rightarrow P$ leakage in the
real data and should be insensitive to systematics in the beam
measurement; $\rho$ estimates are shown in Table~\ref{tab:bmsim_rho}.
Uncertainties in the cross spectra arise from several sources:
instrumental and atmospheric noise, the true-sky lensing and dust $B$
modes, and noise in the beam map measurement.  To estimate the
uncertainty from the CMB half of the cross spectrum, we use our
ensemble of 499 \bmode\ lensing, dust, and noise (i.e. instrumental
and atmospheric) simulations---see BK15 for more details.  Using the
single beam map simulation, we take the 499 cross spectra:
\begin{eqnarray*}
  \left(\mbox{Fixed } T \rightarrow P \right) \times 
  \left(\mbox{lensing} + \mbox{dust} + \mbox{noise} \right)
\end{eqnarray*}
The variance of these cross spectra constitute the error bars in
Figure~\ref{fig:bmsim_result}, which are propagated through to
uncertainties on the $\rho$ estimates (``Cross spectrum $\sigma$'' in
Table~\ref{tab:bmsim_rho}).  Since the beam map jackknife is
negligible compared to this, uncertainty in the beam measurement is
not included here.

\begin{table}
\begin{center}
\caption{Beam Map Simulation $\rho$ Estimates}    
\begin{tabular}{|l|c|c|c|} 
\hline
\rule[-1ex]{0pt}{3.5ex}  \ \ & 95 GHz & 150 GHz & 220 GHz \\
\hline
Upper Limit & $1.3 \times 10^{-2}$ & $1.0 \times 10^{-2}$
& $8 \times 10^{-3}$ \\
Beam Map Jackknife & $1 \times 10^{-6}$ & $1 \times 10^{-5}$ & $4
\times 10^{-4}$ \\
BK15 Cross Spectrum & $4 \times 10^{-3}$ & $5 \times 10^{-3}$  & $2.4
\times 10^{-2}$  \\
Cross Spectrum $\sigma$ & $4 \times 10^{-3}$ & $4 \times 10^{-3}$ &
$1.4 \times 10^{-2}$ \\
\hline
\end{tabular}
\tablecomments{Single-frequency $\rho$ estimates (i.e. equivalent $r$
  level) for the beam map simulation auto spectra (``Upper Limit''), a
  probe of the beam map uncertainty (``Beam Map Jackknife''), the
  cross spectra with real BK15 maps, and uncertainty in the cross
  spectrum arising from the CMB map (``Cross spectrum $\sigma$'').}
\label{tab:bmsim_rho}
\end{center}
\end{table}

\subsection{Discussion}
\label{ss:bmsim_discuss}

Tracing $T \rightarrow P$ leakage from input beam maps to the final
$BB$ spectra provides insight into detector and optics fabrication
effects that can be improved.  For example, the large tile corrugation
feature in the 95\,GHz pixels (Figure~\ref{fig:diffbeams}) impacted
about half of the detectors contributing to the map, and has been
corrected in focal planes produced subsequently.  The contribution of
these detectors is extremely anomalous; our present 95\,GHz $\rho$
results are over an order of magnitude larger than what they would be
if the tile edge detectors were excluded \citep{karkare17}.

Do the beam map simulations correlate with the real BK15 maps and
detect $T \rightarrow P$ leakage?  At 95 and 150\,GHz the cross
spectra are generally positive and follow the auto spectra, but are a
factor of $\sim 2$ lower.  Given the error estimates---$\rho = (4 \pm
4) \times 10^{-3}$ at 95\,GHz and $\rho = (5 \pm 4) \times 10^{-3}$ at
150\,GHz---there is only tentative evidence for leakage in the real
BK15 maps.  The suppression of the cross with respect to the auto
could indicate nonnegligible systematics in the beam map measurement,
which would bias the upper limits high and fail to correlate with the
real maps.

At 220\,GHz there is more power in the real map (due to dust and
higher noise levels; see BK-X), so the large fluctuations in the cross
with real are not unexpected.  It seems likely that the uncertainties
are underestimated.  While Table~\ref{tab:bmsim_rho} indicates that
the beam map statistical errors are a factor of 20 lower than the auto
spectrum, it is possible that systematics in the beam map measurement
contribute extra variance to the cross spectrum that is not captured
in the jackknife estimate.  For example, visual inspection of the beam
maps shows that contamination from the ground and mast is much more
prominent at 220\,GHz than at 95 and 150\,GHz, and may leak into
pixels that are not spatially masked at a level below the noise of an
individual map.  Future work with more rigorous masking of fixed
structure is likely to improve the quality of these maps.  Given these
caveats, the recovered $\rho = (2.4 \pm 1.4) \times 10^{-2}$ should
not be taken as evidence for leakage.

Taken together, the beam map simulations do not definitively detect $T
\rightarrow P$ leakage in cross-correlation with CMB.  The 95 and
150\, GHz maps show a $\sim 1 \sigma$ excess which cannot exclude zero
leakage.  At 220\,GHz the cross spectra fluctuate enough to suggest
underestimated error bars, and no real indication of leakage within
the uncertainty.  We also note that because this frequency mostly
measures dust emission, the sensitivity of $r$ recovery on its leakage
is weak.

\section{The Effect of Undeprojected Residuals on Parameter Recovery}
\label{sec:likelihood}

The \bicep/\keck\ likelihood analysis uses maps at several frequencies
to separate the CMB from foregrounds (see BK-X).  The single-frequency
$\rho$ estimators of $T \rightarrow P$ leakage that we used in
Section~\ref{sec:ttop} are therefore not representative of the
effective bias on $r$ that we may incur from this potential additive
systematic once all spectra are accounted for.  In this Section, we
discuss several methods of dealing with systematics in analysis.
Using the auto and cross spectra presented in Section~\ref{ss:bmsim},
we then simulate the potential effect of this leakage on $r$ recovery
in the multicomponent likelihood analysis.

\subsection{Treating Systematics in Analysis}
\label{ss:syst_treatment}

How we deal with systematics in analysis depends on the uncertainty in
the systematic's form.  In the case of a well-characterized
effect---i.e. in which the specific morphology of the $T
\rightarrow P$ leakage is known---we would simply subtract the leaked
signal in the time or map domain.

If we had intermediate knowledge of the systematic---such as a
well-characterized amplitude, but not specific form---we would
calculate the leakage bandpowers and debias them from the real data.
The uncertainty on the debias would need to be included, e.g. by
inflating the bandpower covariance matrix.

Finally, if there were substantial uncertainty on both the form and
amplitude---e.g. if the uncertainty on the systematic is comparable to
the estimated level of the systematic itself---there is little
argument for debiasing.  In this case, we would run simulations to
determine whether the likely amplitude is small compared the the
experiment's statistical uncertainty.

In Section~\ref{ss:bmsim_discuss} we analyzed the cross spectra of the
beam map simulations with the BK15 maps and found that the uncertainty
on the predicted $T \rightarrow P$ leakage is comparable to its
amplitude (Table~\ref{tab:bmsim_rho}).  Since we cannot verify the
amplitude of this leakage in the real CMB maps at high confidence, we
will show that when propagated through the multicomponent likelihood,
this potential systematic is subdominant compared to the statistical
errors.

\subsection{Simulating $T \rightarrow P$ Leakage in the Multicomponent Analysis}
\label{ss:multicomp}

Here we use simulations to find the bias on $r$ that the BK15
likelihood analysis would incur if $T \rightarrow P$ leakage levels
consistent with either the beam map auto spectra or the cross spectra
between beam maps and CMB data existed in the real maps.  We also test
the possibility of fitting for and marginalizing over a leakage
template.

The BK15 likelihood analysis uses all available $BB$ auto and cross
spectra between the BK15 95, 150, and 220\,GHz maps and external
\planck\ and \wmap\ maps to generate a joint likelihood of the data
for a particular parametric model.  For details on the likelihood
implementation see \citet{bkp, biceptwoVI, biceptwoX}.

To gauge the impact of $T \rightarrow P$ leakage on $r$ recovery, we
analyze the shift in the maximum-likelihood $r$ value
($r_{\text{ML}}$) for sets of 499~simulated bandpowers that have had
bias added corresponding to one of the $T \rightarrow P$ leakage
estimates.  Note that we calculate $\Delta r$ for the peak of the
multi-dimensional likelihood, not the peak of the marginalized
posterior pdf; we find that this provides a more direct view of biases
in the analysis and is easily extended to non-physical negative values
of $r$ to avoid truncation of the distribution.  Histograms of
$r_{\text{ML}}$ from simulations are used as validation in the
baseline BK15 analysis (Figure~20 of~BK-X), and we use the standard
deviation of that distribution as a measure of experimental
sensitivity, $\sigma(r) = 0.020$ for BK15.

\begin{figure}
\begin{center}
\resizebox{0.95\columnwidth}{!}{\includegraphics{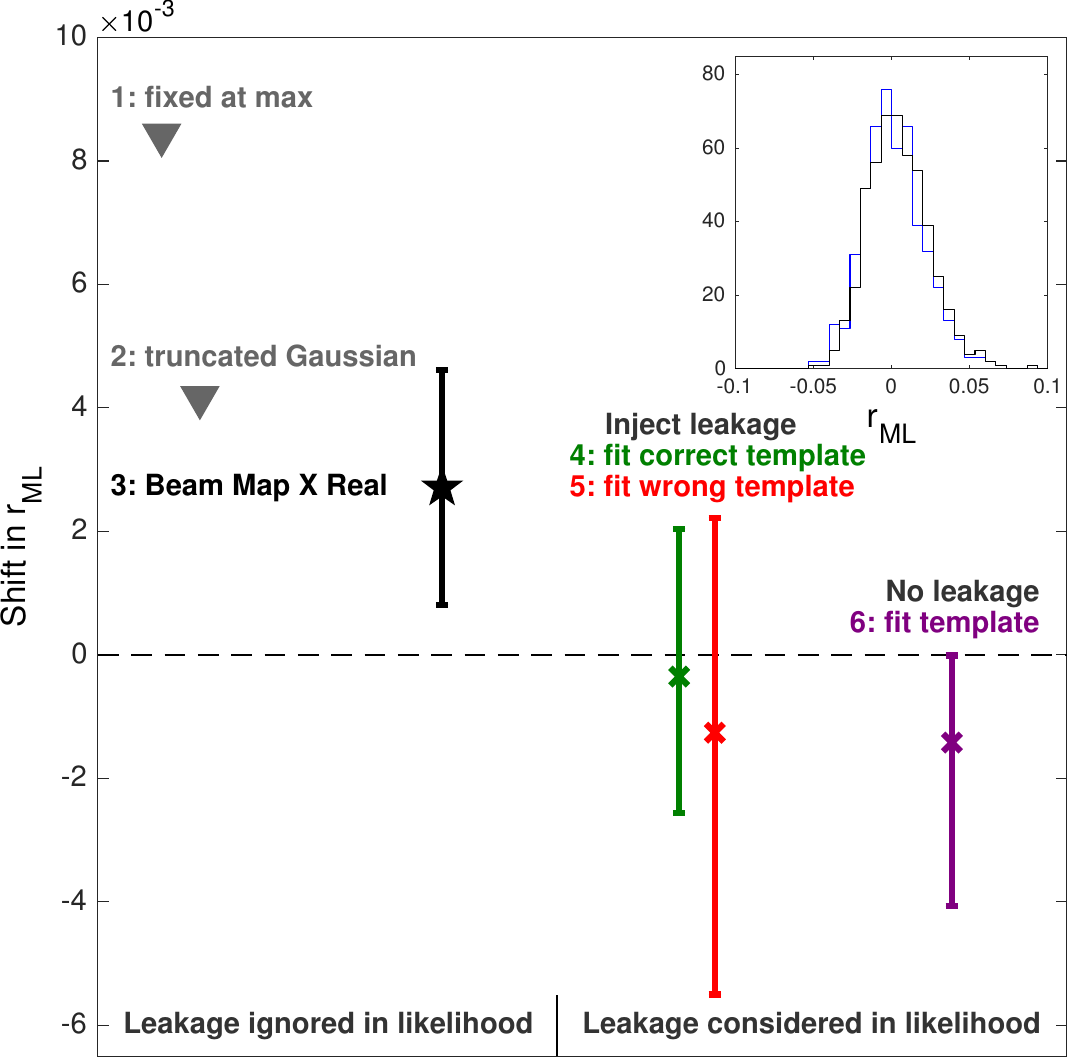}}
\end{center}
\caption{Shifts in maximum-likelihood $r$ with respect to the baseline
  analysis for a set of 499 simulations which have had varying levels
  of $T \rightarrow P$ leakage injected and with various recovery
  scenarios.  The numbers refer to scenarios listed in
  Section~\ref{ss:multicomp}.  For scenarios 3--6, the
  16\textsuperscript{th}, 50\textsuperscript{th}, and
  84\textsuperscript{th} percentiles are plotted.  The inset panel
  shows histograms of $r_{\text{ML}}$ for the baseline analysis (blue) and
  Scenario 3 (black) to emphasize that the potential bias due to $T
  \rightarrow P$ leakage is much smaller than the BK15 statistical
  error, $\sigma(r) = 0.020$.  }
\label{fig:rbias}
\end{figure}

For each $T \rightarrow P$ leakage scenario, we look at the
distribution of realization-by-realization shifts in $r_{\text{ML}}$
relative to a baseline that does not include leakage in the simulation
bandpowers and does not consider leakage parameters in likelihood
analysis (this baseline analysis exactly corresponds to Figure~20
of~BK-X).  For the ``upper limit'' scenarios we report the median of
the 499~shifts, while for the others we report the
16\textsuperscript{th}, 50\textsuperscript{th}, and
84\textsuperscript{th} percentiles to reflect potentially-asymmetric
distributions.  Details of each scenario are listed below; the results
are illustrated in Figure~\ref{fig:rbias}.

\begin{itemize}
  \item Two ``upper limit-driven'' scenarios: We inject leakage
    consistent with the beam map simulation auto spectra (black lines
    in Figure~\ref{fig:bmsim_result}), and ignore it in the likelihood
    analysis.  Recall that the auto spectra are upper limits since
    they may be biased by systematics in the beam measurement.  If we
    simply add the full beam map auto spectra
    (Figure~\ref{fig:bmsim_result} black curves) to the
    simulations---a situation we consider extremely unlikely---we find
    a median bias of $\Delta r = 0.0084$ (Scenario 1).  To better
    reflect our belief in the auto spectra as 95\% upper limits, in a
    second scenario we inject leakage with the same shape as the auto
    spectra, but with variable amplitude drawn from a Gaussian
    distribution centered at zero with $\sigma = 0.5$ of the nominal
    amplitude, and truncated at zero so only positive leakage can be
    added. The median bias is $\Delta r = 0.0042$ (Scenario 2).
  \item A ``CMB data-driven'' scenario: We inject leakage consistent
    with the cross spectra between the beam map simulations and the
    BK15 maps, and ignore it in likelihood analysis.  Specifically,
    each realization is biased by a leakage contribution that is
    randomly drawn from the Figure~\ref{fig:bmsim_result} teal points
    and error bars.  This is an attempt to model the leakage that
    appears to actually exist in the maps, albeit at marginal
    significance.  The recovered bias is $\Delta r = 0.0027 \pm
    0.0019$ (Scenario 3).
  \item Three ``recovery'' scenarios: Using knowledge of the leakage
    from beam map simulations, we attempt to marginalize over it in
    the likelihood.  For each frequency we add a new parameter to the
    likelihood analysis, which scales the leakage bias contribution to
    the bandpowers.  For example, the parameter $\eta_{95} = 0$
    represents zero $T \rightarrow P$ leakage in the 95~GHz map, while
    $\eta_{95} = 1$ indicates leakage equal to that shown in the left
    panel of Figure~\ref{fig:bmsim_result}.  For this analysis, we
    must select one of the two leakage estimates to use as a
    template---either the beam map simulation auto spectrum or the
    beam map simulation cross spectrum with BK15 maps.  We use a flat
    prior on $\eta$ in the range of $[0,2]$.  If the template used in
    the analysis matches the leakage added to the simulations, the
    resulting bias on $r$ is small: $\Delta r =
    -0.0004^{+0.0024}_{-0.0022}$ (Scenario 4).  If the \emph{wrong}
    template is used (e.g. inject the beam map/CMB cross spectrum, but
    fit for the beam map auto spectrum), we incur a negative bias:
    $\Delta r = -0.0013^{+0.0035}_{-0.0042}$ (Scenario 5).  Finally,
    if \emph{no leakage} is injected but we attempt to fit for it, we
    still find negative bias with $\Delta r =
    -0.0014^{+0.0014}_{-0.0027}$ (Scenario 6).

\end{itemize}

We take the ``CMB data-driven'' Scenario 3, $\Delta r = 0.0027 \pm
0.0019$, as our best estimate of the bias on $r$ from $T \rightarrow
P$ leakage.  Figure~\ref{fig:rbias} illustrates that in all cases
the bias is subdominant compared to the BK15 statistical
error, $\sigma(r) = 0.020$.  Quadrature addition of the worst-case
upper limit bias inflates $\sigma(r)$ by 8\%.

In the standard \texttt{COSMOMC} analysis of e.g. BK15, physical
priors are imposed.  If we introduce a component that i) is partially
degenerate with the signal of interest ($r$), ii) is not clearly
detected at the available noise level, and iii) has a positive-only
prior applied, $r$ tends to be biased downwards.  Since we do not
clearly detect leakage in the current data, we decide not to
marginalize over the $\eta$ parameters in the $r$ constraint analysis
at this time.  This is analogous to the treatment of the dust
decorrelation parameter as discussed in Appendix F of BK15.  We will
continuously review this situation going forward.

\section{Conclusions}
\label{sec:conclusion}

In this paper we presented far-field optical characterization of
\keckarray\ detectors at 95, 150, and 220\,GHz contributing to the
BK15 dataset.  From an extensive far-field beam mapping campaign we
measured differential Gaussian parameters with high precision, which
are repeatable from year to year.  For each detector we formed deep
composite maps covering all azimuthal angles, and from them generated
array-averaged beam profiles.

The composite beam maps were then used to predict the residual main
beam $T \rightarrow P$ leakage expected in the BK15 polarization maps
after deprojection of the lowest-order beam mismatch modes.  From an
auto spectrum analysis of beam map simulations, noting that there may
exist contributions from low-level systematics in the beam
measurement, we presented upper limits to the leakage.  Cross spectra
between the beam map simulations and the BK15 CMB maps offer tentative
evidence for leakage, but the uncertainties are large enough that zero
leakage cannot be excluded.

We have run simulations using the BK15 multicomponent likelihood
analysis to test the effect of undeprojected $T \rightarrow P$ leakage
on $r$ recovery.  When leakage consistent with the cross spectra
between beam map simulations and real CMB maps is added, the bias is
$\Delta r = 0.0027 \pm 0.0019$.  It is possible to marginalize over
this contribution in the multicomponent analysis, but this admits the
possibility of a small negative bias if the wrong template is used
with physical priors.  Because the leakage is not clearly detected in
the CMB data, we do not marginalize over it in the current constraint
on $r$. All of the biases presented are small compared to the BK15
statistical uncertainty $\sigma(r) = 0.020$.

\bicep's sensitivity to $r$ will continue to steadily improve: with
data taken through the 2017 season we expect $\sigma(r) \sim 0.010$,
and with the \bicep\ Array experiment under construction we anticipate
$\sigma(r) < 0.005$ within five years \citep{hui18}.  Constraints on
beam systematics will need to similarly tighten.  Future effort will
focus on three aspects of the problem: the intrinsic $T \rightarrow P$
leakage level, the measurement thereof, and treatment in analysis.

High-fidelity beam maps point to features in individual beams that
contribute significant $T \rightarrow P$ leakage, which can then be
remedied in hardware.  Such feedback has been critical to ensuring
that leakage is reduced in later generations of receivers.  For
example, the 95\,GHz difference beam shown in
Figure~\ref{fig:diffbeams} highlighted an anomalous interaction
between the tile corrugations and edge detectors.  Compared to
detectors that were not affected, the tile edge detectors drove up the
estimated $T \rightarrow P$ leakage for the BK15 results by over an
order of magnitude.  We corrected this effect in focal planes produced
subsequently, and expect that this feedback cycle will continue.

Looking beyond the BK15 result, we have already improved the beam map
reduction compared to the maps presented in this paper: non-Gaussian
noise has been significantly reduced through more optimal low-level
deglitching and demodulation.  Re-reduction of existing data will
produce cleaner maps and allow us to mean-filter the component maps
when generating the composites.  In future beam mapping campaigns we
also plan to modify the raster scan strategy to produce robust noise
estimates, e.g. by using out-and-back scans at the same elevation to
form a ``scan direction'' jackknife.  Maps generated from these data
will be used to generate beam map noise realizations without dividing
the composite maps into two halves.  Their enhanced statistical
properties will facilitate comparison of $T \rightarrow P$ leakage to
CMB data.

While in this paper we have shown a basic attempt to detect $T
\rightarrow P$ leakage by cross-correlating the final coadded leakage
and real CMB maps (in which much of the leakage has cancelled), this
comparison can be improved.  In our next results we plan to form cross
spectra that detect the predicted leakage at high significance.  We
will isolate high-leakage detector subsets or form combinations of
on-sky data that are expected to enhance the leakage (e.g. boresight
angles that do not cancel) in particular sub-maps compared to the full
dataset.  Given these higher-confidence estimators of $T \rightarrow
P$ leakage in the CMB maps, we will explore several options to remove
the effect.  If the number of high-leakage detectors is small,
excluding or de-weighting them will dramatically improve systematic
control.  With improved statistical properties of the beam maps,
debiasing as discussed in Section~\ref{ss:syst_treatment} would also
be reasonable.  Finally, we can further reduce $T \rightarrow P$
residuals by deprojecting additional modes of this leakage for each
detector pair.  This can be done using modes drawn from bases
independent of the beam maps, by using the beam maps to directly
predict the template of each pair's undeprojected residual, or by
using these maps to guide definition of a small subset of modes to be
deprojected.

\acknowledgments

The \keckarray\ project has been made possible through
support from the National Science Foundation under Grants
ANT-1145172 (Harvard), ANT-1145143 (Minnesota)
\& ANT-1145248 (Stanford), and from the Keck Foundation
(Caltech).
The development of antenna-coupled detector technology was supported
by the JPL Research and Technology Development Fund and Grants No.\
06-ARPA206-0040 and 10-SAT10-0017 from the NASA APRA and SAT programs.
The development and testing of focal planes were supported
by the Gordon and Betty Moore Foundation at Caltech.
Readout electronics were supported by a Canada Foundation
for Innovation grant to UBC.
The computations in this paper were run on the Odyssey cluster
supported by the FAS Science Division Research Computing Group at
Harvard University.
The analysis effort at Stanford and SLAC is partially supported by
the U.S. DoE Office of Science.
We thank the staff of the U.S. Antarctic Program and in particular
the South Pole Station without whose help this research would not
have been possible.
Most special thanks go to our heroic winter-overs Robert Schwarz
and Steffen Richter.
We thank all those who have contributed past efforts to the \bicep--\keckarray\
series of experiments, including the \bicepone\ team.
We also thank the \planck\ and \wmap\ teams for the use of their
data.

\bibliography{ms}

\end{document}

%% file: defs.tex
\def\bicep{BICEP}

\def\bicepone{{\sc BICEP1}}
\def\biceptwo{{\sc BICEP2}}
\def\bicepthree{{\sc BICEP3}}

\def\keck{{\it Keck}}
\def\keckarray{{\it Keck Array}}

\def\planck{{\it Planck}} 

\def\wmap{WMAP}

\def\healpix{{\sc healp}ix}

\def\bmode{$B$-mode}

\def\lcdm{$\Lambda$CDM}

\def\aap{Astron.\ Astrophys.}

\def\apjl{Astrophys.\ J.\ Lett.}
\def\apjs{Astrophys.\ J.\ Suppl.\ Ser.}